\documentclass[twocolumn,showpacs,prx,aps,longbibliography,superscriptaddress]{revtex4-1}
\usepackage[bookmarks=false,linkcolor=blue,urlcolor=blue,colorlinks,citecolor=blue]{hyperref}
\usepackage{tabularray}
\pdfoutput=1

\usepackage{xcolor}
\usepackage{float}
\usepackage{svg}
\usepackage{adjustbox}
\usepackage{graphicx}
\usepackage{color}
\usepackage{amsmath}
\usepackage{amsfonts}
\usepackage{mathrsfs}
\usepackage{amsthm}
\usepackage{amssymb}
\usepackage{wrapfig}
\usepackage[shortlabels]{enumitem}
\usepackage[retainorgcmds]{IEEEtrantools}
\usepackage{tcolorbox}
\usepackage{enumitem}

\makeatletter
\AtBeginDocument{\let\LS@rot\@undefined}
\makeatother

\begin{document}
	\author{Lior Oppenheim}
	\affiliation{The Racah Institute of Physics, The Hebrew University of Jerusalem, Jerusalem 9190401, Israel}
	\author{Maciej Koch-Janusz}
	\affiliation{Department of Physics, University of Zurich, 8057 Zurich, Switzerland}
	\affiliation{James Franck Institute, The University of Chicago, Chicago, Illinois 60637, USA}
    \affiliation{Haiqu Inc., 95 Third Street, San Francisco, California 94103, USA}
	\author{Snir Gazit}
\affiliation{The Racah Institute of Physics, The Hebrew University of Jerusalem, Jerusalem 9190401, Israel}
\affiliation{The Fritz Haber Research Center for Molecular Dynamics, The Hebrew University of Jerusalem, Jerusalem 91904, Israel}
	\author{Zohar Ringel}
\affiliation{The Racah Institute of Physics, The Hebrew University of Jerusalem, Jerusalem 9190401, Israel}
        \title{Machine Learning the Operator Content of the Critical Self-Dual Ising-Higgs Gauge Model}
	
\begin{abstract}
We study the critical properties of the Ising-Higgs gauge theory in $(2+1)D$ along the self-dual line which have recently been a subject of debate. For the first time, using machine learning techniques, we determine the low energy operator content of the associated field theory. Our approach enables us to largely refute the existence of an emergent current operator and with it the standing conjecture that this transition is of the $XY^*$ universality class. We contrast these results with the ones obtained for the $(2+1)D$ Ashkin-Teller transverse field Ising model where we find the expected current operator. Our numerical technique extends the recently proposed Real-Space Mutual Information allowing us to extract sub-leading non-linear operators. This allows a controlled and computationally scalable approach to target the CFT spectrum and discern universality classes beyond $(1+1)D$ from Monte Carlo data.  
\end{abstract}
	
\maketitle
 
\textit{Introduction - }The hallmark of critical phenomena is the emergence of a universal behavior governing the long wave-length theory. In this limit, dynamics are often controlled by collective degrees of freedom dictated solely by symmetry and dimensionality. A case in point is symmetry-breaking transitions, whose critical fluctuations are governed by an order parameter directed along the symmetry-breaking axis \cite{landau_1980}. 

A major challenge in modern condensed matter theory is addressing critical phenomena beyond the above Landau paradigm \cite{sachdev_2010,xu_2012}. This includes spin liquids \cite{xiao_2007}, fractional Hall effect \cite{girvin_2005}, and symmetry protected topological transitions \cite{barkeshli_2016,barkeshli_2019}. In such cases, identifying the low-energy theory often becomes a formidable task, due to the absence of clear symmetry-based candidates for the low-lying degrees of freedom \cite{senthil_2004,sachdev_2019,sachdev_2023,gazit_2023}. 

Lattice gauge theories provide a paradigm for studying such criticalities. In this setting, the transition is described by a condensation of gauge field fluxes, charged matter fields, or both. The former is known as the confinement transition, and the latter as the Higgs transition. In their seminal work, Fradkin and Shankar provided a unified framework describing both transitions \cite{fradkin_1979}. Nevertheless, the case in which both transitions occur simultaneously at a multicritical point (MCP) remained elusive. 

This outstanding problem attracted much recent interest, particularly the case of the Self-Dual Ising-Higgs Gauge theory in $(2+1)D$ (``SD-IHG")\cite{kitaev_2003,vidal_2009,tupitsyn_2010,gazit_2018,nahum_2021,iqbal_2022,bonati_2022_0,bonati_2022,manoj_2023}. Here the two transitions meeting at the MCP are of the $3D$ Ising and Ising* universality classes. Following the general insight that MCPs of two Ising transitions may lead to an enhanced $U(1)$ symmetry Bonati \emph{et al.~}\cite{bonati_2022} argued, with supporting Monte Carlo results, that the emergent theory is of the $XY^*$ type \cite{isakov_2012} --- an $XY$ transition exhibiting only gauge invariant operators. However, in the absence of a direct identification of low energy degrees of freedom in terms of the microscopic ones, the validity of such a phenomenological description is unclear.

The ideal way to indisputably verify this conjecture is to obtain the operator spectrum of the theory, or at least its leading orders. Indeed, the putative $XY^*$ transition should contain a smoking gun: three degenerate operators with scaling dimension $2$, namely the three vector components of the current operator associated with the emergent $U(1)$ symmetry. While for $2D$ critical points, such data is readily accessible through transfer matrix diagonalization \cite{cardy_1986}, in $(2+1)D/3D$ it is a challenging numerical problem. Despite recent progress \cite{schuler_2022,zhou_2023,zhu_2023}, we currently lack a generic tool for this task.  

More broadly, extracting the operator content beyond the leading order from microscopic samples, thus constructively connecting the micro- and macroscopic descriptions, is an open challenge in many fields. Recently, methods based on information theory and deep learning have shown promise in this task \cite{ringel_2018,li_2018,sante_2022,efe2_2021,margalit_2022,zhang_2023}. One approach, supported by analytical guarantees \cite{gordon_2021}, is the Real-Space Mutual Information Neural Estimator (RSMI-NE) algorithm, which was used to identify and extract leading operators in the field theory from microscopic Monte Carlo samples \cite{efe2_2021,efe_2021,efe_2023}. The possibility of using such techniques to methodically extract sub-leading parts of the operator spectrum remained, however, unexplored. 

In this work, we address the question of the MCP in the SD-IHG theory in $(2+1)D$ numerically.
To this end, we extend the RSMI-NE algorithm, enabling a systematic extraction of sub-leading orders of operators in the spectrum. Applying this technique to the model of interest, we obtain both leading and sub-leading operators, namely the energy operator and its derivatives. Crucially, a current operator does {\emph not} appear in the spectrum. This is in contrast to a model known to exhibit an emergent $U(1)$ symmetry based on coinciding Ising transitions, where we obtain all expected operators, \emph{up to and including the current operator}. We thus rule out the existence of a local current operator for the SD-IHG theory and, with it, the classification of the critical theory as $XY^*$.

\textit{Models - }We investigate two models: the SD-IHG model -- the principal subject of interest -- and the Ashkin-Teller transverse field Ising (AT-TFI) theory in $(2+1)D$, which is used to compare and contrast the numerical results \cite{schuler_2022}.

\begin{figure}[htp]
\centering
\adjustbox{trim={0.03\width} {.45\height} {0.2\width} {0\height},clip}{
\includegraphics[width=0.62\textwidth]{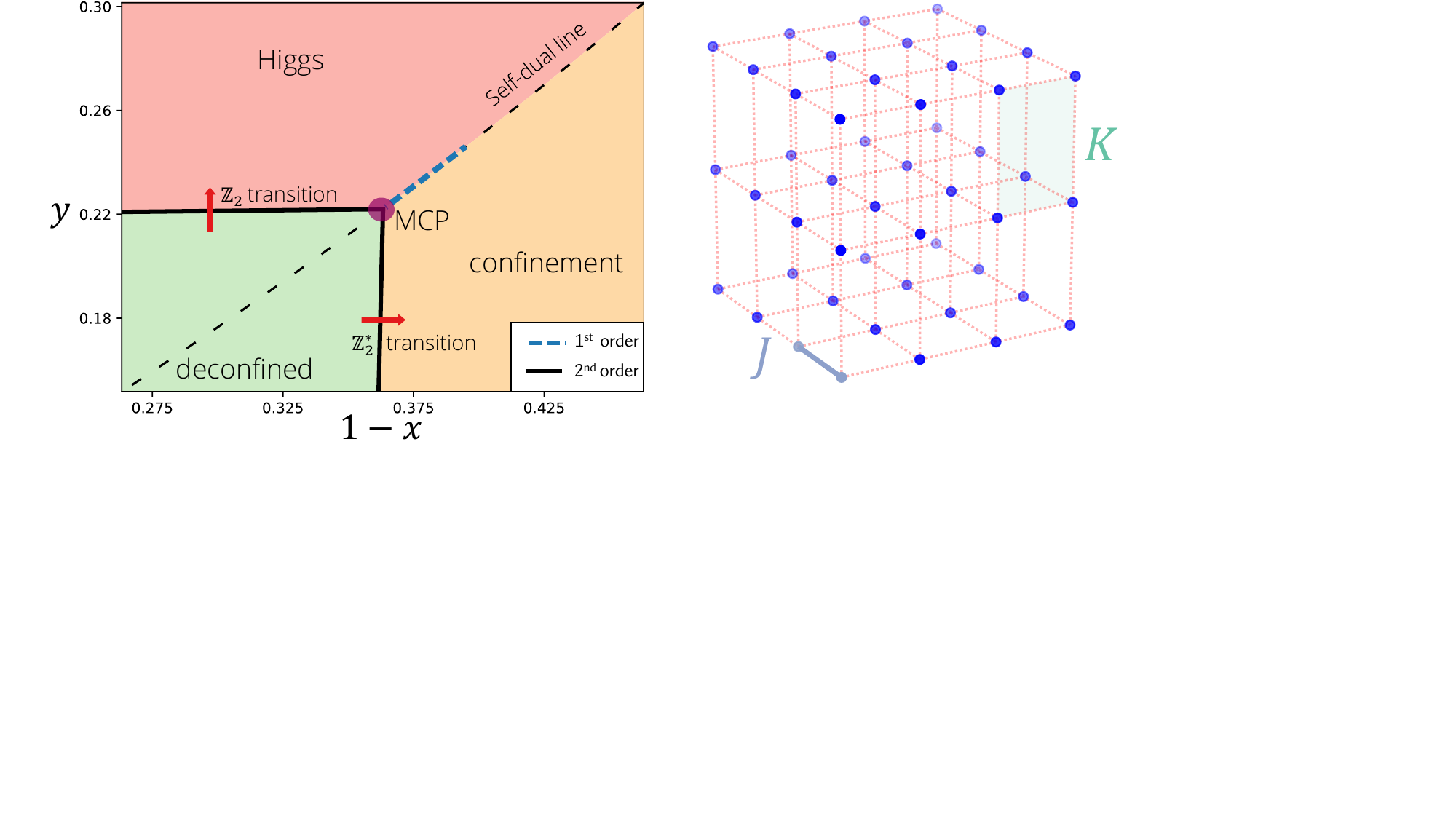}}
\caption{\textit{(Left)} Phase diagram for the self-dual Ising-Higgs gauge model. The two Ising second-order transitions meet at the MCP, which is conjectured to belong to the $XY^\ast$ universality class. A first-order transition also takes place along the self-dual line. \textit{(Right)} Visual representation of the action $\mathcal{S}_{\text{\tiny SD-IHG}}$ (Eq. \ref{eq:H_SDIHG}). The blue sites represent the matter field $\tau_i$ and the red bonds represent the gauge field $\sigma_{ij}$}
    \label{fig:phase_and_H_SDIHG}
\end{figure}

The classical SD-IHG model describes $\mathbb{Z}_2$ gauge- and matter- fields $\sigma_{ij}=\pm 1$ and $\tau_i=\pm 1$, residing, respectively, on the bonds and sites of a cubic $(2+1)D$ lattice. The space-time action is given by (see Fig. \ref{fig:phase_and_H_SDIHG}):
\begin{equation}\label{eq:H_SDIHG}
\mathcal{S}_{\text{\tiny SD-IHG}}=K\sum_\square\prod_{\left<i,j\right>\in\square}\sigma_{ij}+J\sum_{\left<i,j\right>}\tau_{i}\sigma_{ij}\tau_{j}
\end{equation}
The first term describes the interaction of four gauge fields around a shared plaquette, while the second describes the interaction of two adjacent matter fields, which is mediated by the gauge field. The gauge symmetry is manifested in the local transformations $\sigma_{ij}\to \eta_{i}\eta_{j}\sigma_{ij}$ and $\tau_i\to \eta_i \tau_i$ (with $\eta_i=\pm 1$), under which the action remains invariant. The gauge-invariant quantities are either a closed loop of gauge fields (Wilson loop) or two matter fields with a string of gauge fields stretched between them (Wilson string).

The model admits a duality mapping relating the parameters as follows:
$$x'=\frac{1-y}{1+y},\ \ y'=\frac{1-x}{1+x},$$
where $x=\mathrm{tanh}(K)$ and $y=\mathrm{tanh}(J)$. The system is self-dual along the line $x=\frac{1-y}{1+y}$. In the extreme case of $x\to 1$, the gauge fields are stiff, and the theory reduces to that of an Ising model, with an Ising transition at $y_c\approx 0.218$. In the other extreme case, $y\to 0$, the theory reduces to that of a pure $\mathbb{Z}_2$ gauge model, with an Ising$^*$ transition at $x_c\approx 0.642$. These two transitions are stable even for finite $K$ ($x<1$) or a non-zero $J$ ($y>0$) \cite{fradkin_1979}. Their meeting point on the self-dual line forms an MCP at $x_{\text{\tiny MCP}}\approx 0.6367$ \cite{nahum_2021} (See Fig. \ref{fig:phase_and_H_SDIHG}).

While the complete nature of the multi-criticality for the SD-IHG model is unknown, the self-duality symmetry allows us to infer the exact form of two relevant primary operators in the spectrum which are symmetric and anti-symmetric under its action ($S$ and $A$ operators)\cite{nahum_2021}:
\begin{align}
  \label{eq:AS}
  \begin{split}
&  A = \langle B\rangle +\frac{2x_{\text{\tiny MCP}}}{1-x_{\text{\tiny MCP}}^2}\langle P\rangle -\frac{1}{1-x_{\text{\tiny MCP}}} \\
&  S =\langle B\rangle -\frac{2x_{\text{\tiny MCP}}}{1-x_{\text{\tiny MCP}}^2}\langle P\rangle +\frac{1}{1-x_{\text{\tiny MCP}}}
  \end{split}
\end{align}
Where $\langle P\rangle$ and $\langle B\rangle$ are spatial averages of the six plaquettes and twelve bonds of a cube.

The second system we consider is the AT-TFI model in $(2+1)D$, which serves as a benchmark of our approach, as it contains a fully understood, yet non-trivial, critical point of the $XY^*$ universality class, described by a similar field theory to the one proposed in \cite{bonati_2022}.

\begin{figure}[htp]
\centering
\adjustbox{trim={0.00\width} {.4\height} {0.2\width} {0\height},clip}{
\includegraphics[width=0.58\textwidth]{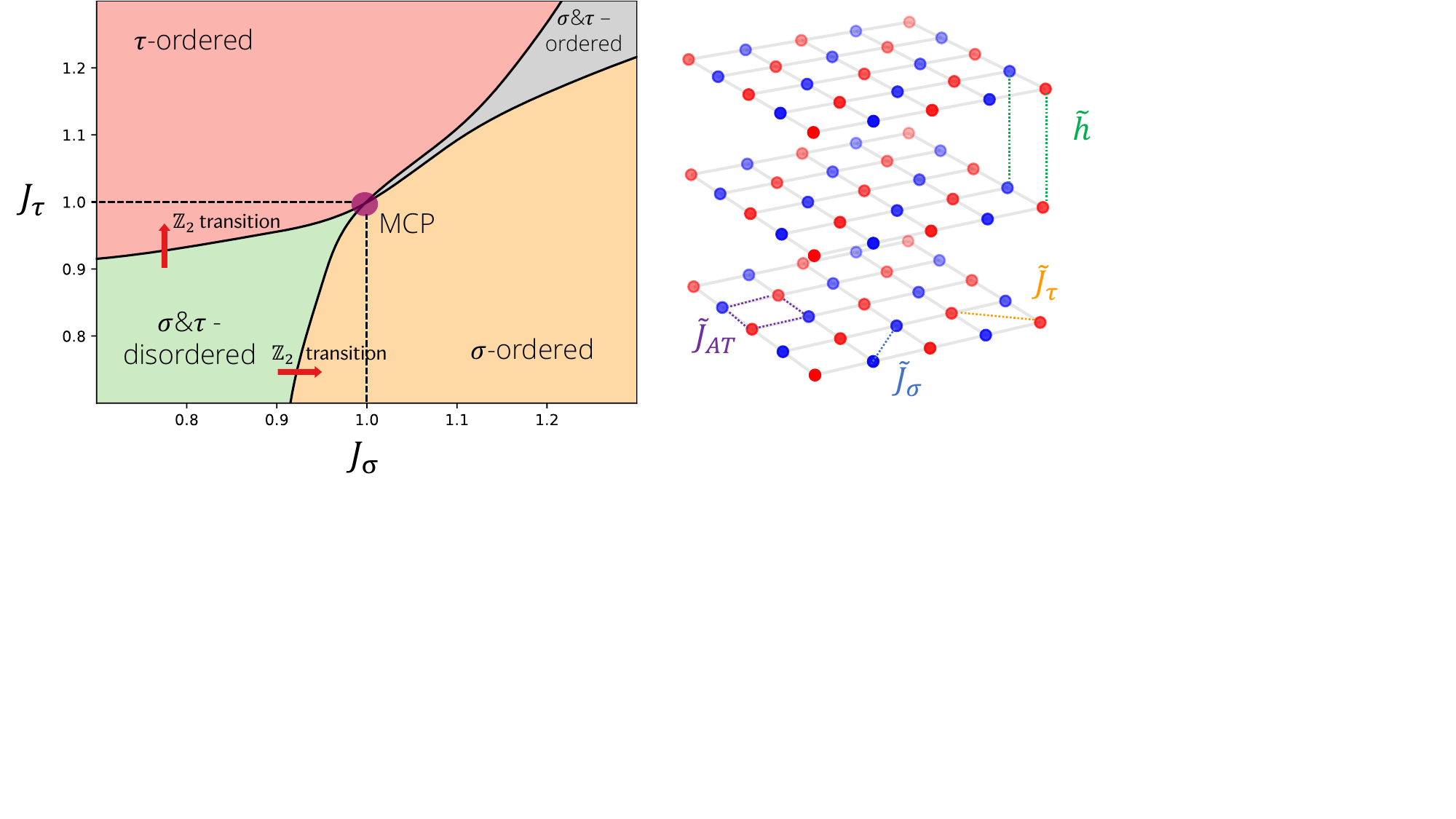}}
\caption{\textit{(Left)} Phase diagram of the Ashkin-Teller Transverse-Field Ising (AT-TFI) model in $(2+1)D$, as a function of the Ising couplings $J_\sigma$ and $J_\tau$, with $J_{AT}$ and $h$ tuned to criticality. The four phases correspond to the separate ordering of the two Ising fields. The two Ising transitions meet at the MCP when $J_\sigma=J_\tau=1$. Due to similarities between this model and the SD-IHG model, we use it as a benchmark for our method \textit{(Right)} Visual representation of $\hat{\mathcal{H}}_{AT-TFI}$ (Eq. \ref{eq:H_ATTFI})  in imaginary time (see appendix \ref{sec:monte_carlo_ATTFI}), showcasing the interlaced sub-lattices form of the corresponding classical model and different interaction terms. The blue and red sites are the $\sigma$ and $\tau$ fields. The $\tilde{J}_{\sigma},\tilde{J}_{\tau},\tilde{J}_{AT},\tilde{h}$ are the coupling constants after the quantum-to-classical mapping.}
    \label{fig:phase_and_H_ATTFI}
\end{figure}

The model is phrased in terms of quantum spins $\hat{\sigma}_i,\hat{\tau}_i$ residing on interlaced sublattices of a $2D$ square lattice. The Hamiltonian is given by \cite{schuler_2022} (See Fig. \ref{fig:phase_and_H_ATTFI}):
\begin{align}
  \label{eq:H_ATTFI}
  \begin{split}
    \hat{\mathcal{H}}_{\text{\tiny AT-TFI}}= & -h\sum_{i}\hat{\sigma}_{i}^{x}+\hat{\tau}_{i}^{x} \\
    & -\sum_{\left\langle i,j\right\rangle}\left[J_{\sigma}\hat{\sigma}_{i}^{z}\hat{\sigma}_{j}^{z}+J_{\tau}\hat{\tau}_{i}^{z}\hat{\tau}_{j}^{z}-J_{AT}\hat{\sigma}_{i}^{z}\hat{\sigma}_{j}^{z}\hat{\tau}_{i}^{z}\hat{\tau}_{j}^{z}\right] 
  \end{split}
\end{align}
and comprises two transverse-field Ising models (TFIM), which reside on interlacing sub-lattices, with a quartic bond-bond interaction. For $J_{AT}>0$ and by setting $h$ to criticality, the model exhibits a quantum $XY$ phase transition at $J_{\sigma}=J_{\tau}=1$. At the critical point, the two Ising transitions meet, and the discrete $\mathbb{Z}_2\times\mathbb{Z}_2$ symmetry is promoted to a continuous $U(1)$ symmetry. At that criticality, the $\sigma$ and $\tau$ field merge into a continuous complex field $\psi=\sigma+i\tau$ which forms a $\left| \psi \right|^4$ theory \cite{schuler_2022}.

As such, the relevant part of the CFT spectrum of the theory is completely characterized. It consists of three primary operators: charge-0, charge-2 (twice degenerate), and a Noether current (thrice degenerate). 

\textit{Methods - }Recently a correspondence between the solutions to a certain mutual-information-based variational problem and the leading operators/eigenvectors in the transfer matrix spectrum was shown \cite{gordon_2021}. This result explains how eigenvectors and eigenvalues of the transfer matrix can be learned using the framework of the information-bottleneck (IB) compression theory \cite{tishbi_1999}. Together with the progress in mutual information estimation algorithms \cite{oord_2019,mine_2018}, this allows us to cast the problem of computing leading and sub-leading eigenvalues and eigenvectors to an unsupervised machine learning problem \cite{efe_2021}.

To leverage this theoretical development we use (and extend) the Real-Space Mutual-Information Neural Estimator (RSMI-NE) algorithm \cite{efe2_2021,efe_2021}. The latter constructs relevant local degrees of freedom based solely on a corpus of Monte Carlo samples of the system under investigation. The input to the algorithm are pairs of random variables $(\mathscr{V},\mathscr{E}$) where $\mathscr{V}$ is a spatial block of the system and $\mathscr{E}$ is a distant environment of $\mathscr{V}$, spatially separated from it by a buffer $\mathscr{B}$. The output of the algorithm is an ordered set of encoders parameterized by neural networks (``neural operators"), which take a configuration $v\in\mathscr{V}$ and compute the values, in decreasing order of relevancy, of the primary and descendant operators in the CFT spectrum (including degeneracies) acting on $v$. 

\begin{table*}[htp]
 \caption{AT-TFI and SD-IHG leading operators content}
\begin{tblr}{
width=1.0\linewidth,
colspec = {X[c]X[c]X[c,h]X[c,h]X[c]X[c]X[c,h]X[c,h]},
stretch = 0,
rowsep = 2pt,
hlines = {black, 1pt},
vlines = {black, 0pt},
vline{5}={2pt},
row{1} = {0.6cm},
column{2} = {1.6cm},
column{6} = {1.2cm},
cell{1}{1,5} = {c=4}{c},
cell{2}{3,7} = {c=2}{c},
cell{2}{1,2,5,6} = {r=2}{c}
}
     AT-TFI  & & & & SD-IHG   & & &  \\
    \SetCell[c=1]{c}{\scriptsize{RSMI-NE Scaling Dimension} \\ \scriptsize \{Expected\textsuperscript{\citep{chester_2019}}\}} &  \SetCell[c=1]{c}{\scriptsize{Analytic Operator \\ \{Deg.\}}}  & \scriptsize{Neural Operator Projection}  & & \SetCell[c=1]{c}{\scriptsize{RSMI-NE Scaling Dimension} \\ \scriptsize \{Expected\textsuperscript{\citep{nahum_2021}}\}}  &  \SetCell[c=1]{c}{\scriptsize{Analytic Operator \\ \{Deg.\}}}  &   \scriptsize{Neural Operator Projection} &  \\
    & & \scriptsize Maximum & \scriptsize Minimum & & & \scriptsize Maximum &  \scriptsize Minimum \\
    \SetCell[c=1]{m}{
  \vbox{\small\( \displaystyle 1.24(1)\)} \\
  \vbox{\small\( \displaystyle 1.22(1)\)}
  \vbox{\small\( \displaystyle \{1.23629\}\)}} & \SetCell[c=1]{m}{\mbox{\small\( \displaystyle \langle \sigma \rangle^2-\langle \tau \rangle^2\)}\\\mbox{\small\( \displaystyle \langle\sigma\rangle\langle\tau\rangle\)} \\  \mbox{\small \{2\}} } &
  \adjustbox{trim={0.00\width} {0.05\height} {0.45\width} {0.00\height},clip}{
  \includegraphics[width=0.12\textwidth]{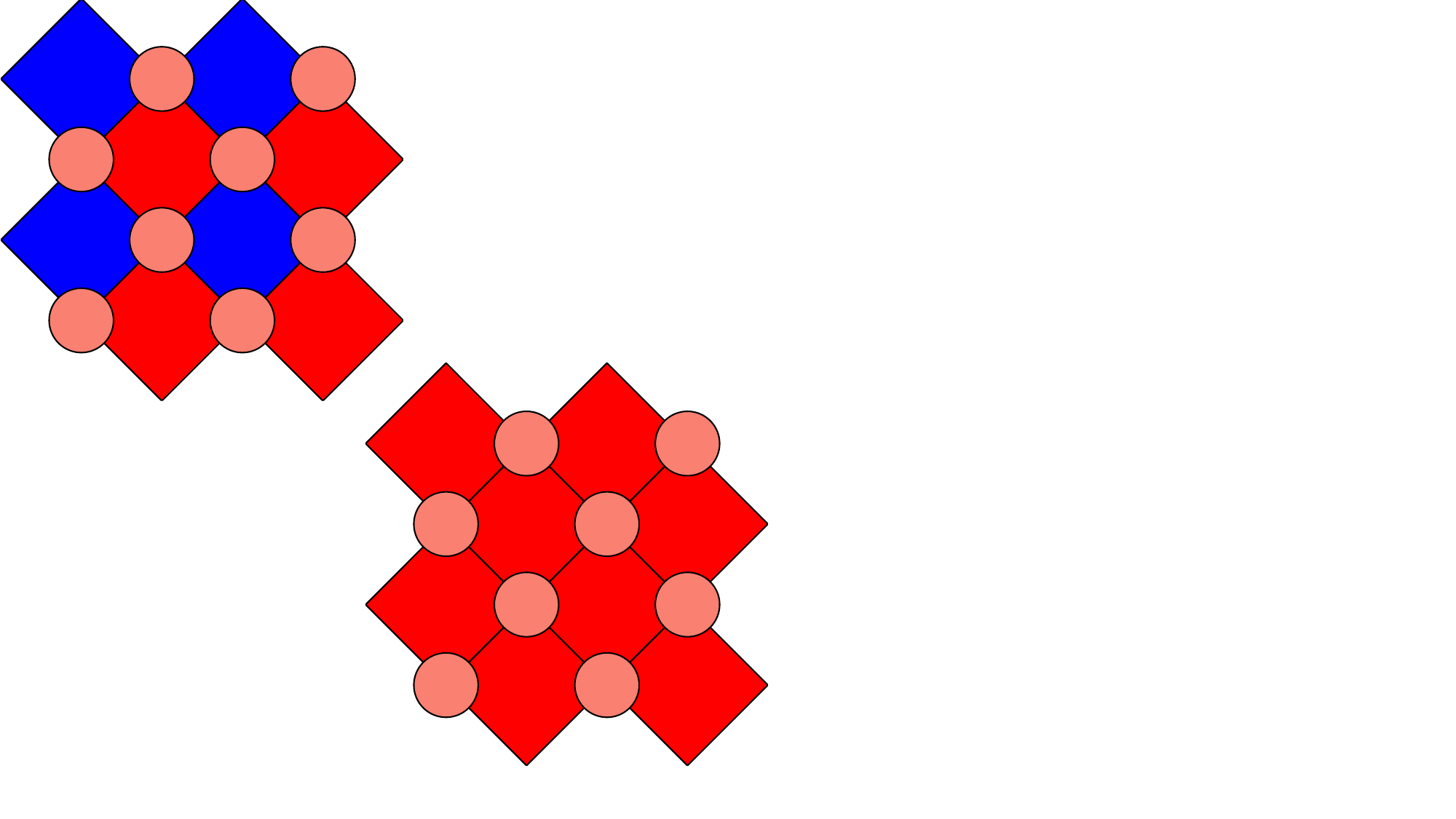}} &
 \adjustbox{trim={0.00\width} {0.05\height} {0.45\width} {0.00\height},clip}{
  \includegraphics[width=0.12\textwidth]{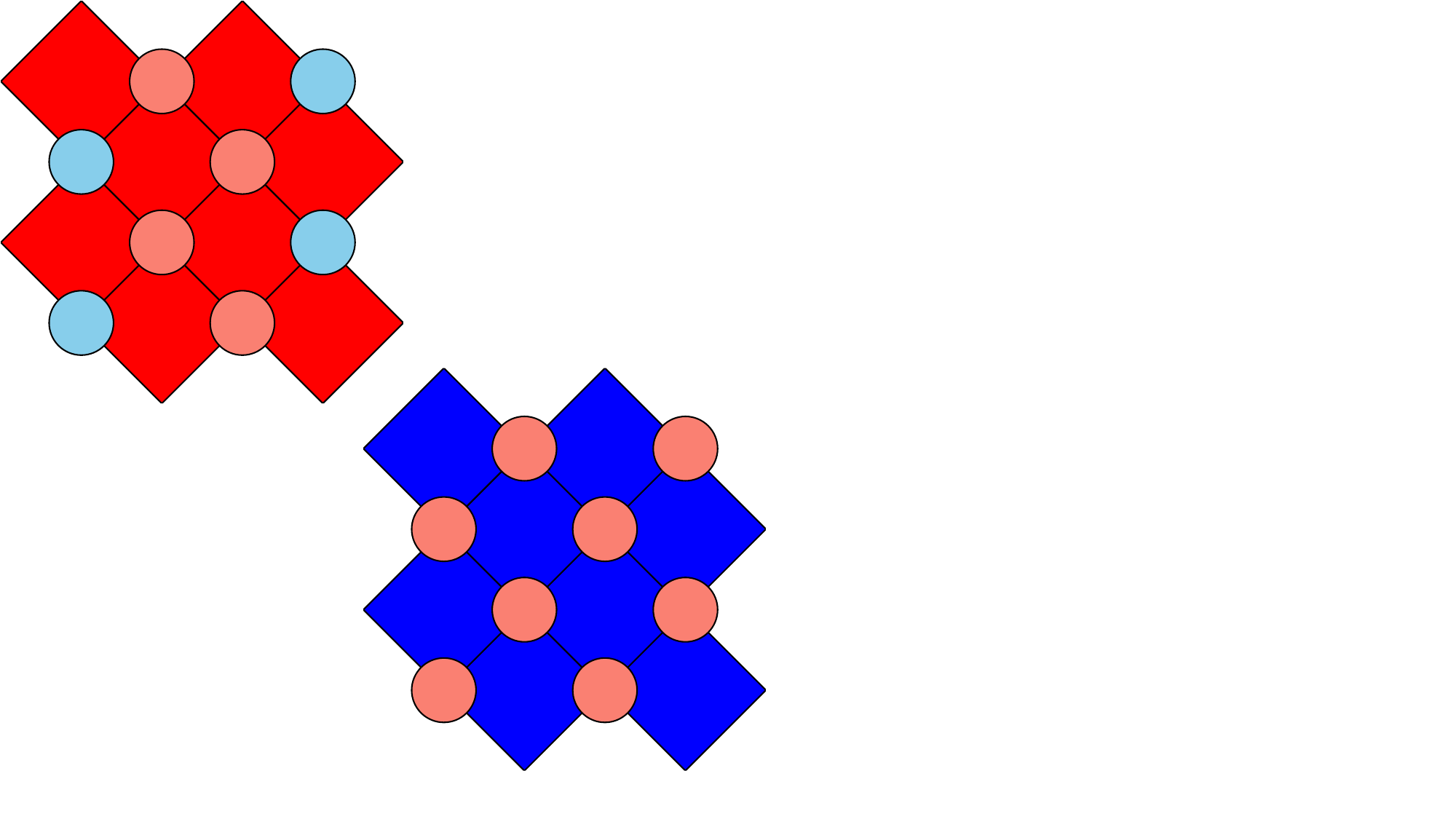}}  & 
    \SetCell[c=1]{m}{
  \vbox{\small\( \displaystyle\ 1.24(1)\)} \\
  \vbox{\small\( \displaystyle \{1.222\}\)}} & \SetCell[c=1]{m}{\vbox{\small\( \displaystyle \langle A\rangle\)} \\ \mbox{\small \{1\}}} &
  \includesvg[width=0.06\textwidth]{A_min.svg} &  \includesvg[width=0.06\textwidth]{A_max.svg} 
    \\
        \SetCell[c=1]{c}{
  \mbox{\small\( \displaystyle 1.49(2)\)} \\
  \mbox{\small\( \displaystyle \{1.51136\}\)}} & \SetCell[c=1]{c}{\mbox{\small\( \displaystyle \langle \sigma\rangle^2+\langle\tau\rangle^2\)} \\ \mbox{\small \{1\}} } &
  \includesvg[width=0.06\textwidth]{epsilon_min.svg} &
  \includesvg[width=0.06\textwidth]{epsilon_max.svg} & 
    \SetCell[c=1]{c}{
  \mbox{\small\( \displaystyle 1.54(2)\)} \\
  \mbox{\small\( \displaystyle \{1.502\}\)}} & \SetCell[c=1]{c}{ \mbox{\small\( \displaystyle \langle S\rangle\)} \\ \mbox{\small \{1\}}} &
  \includesvg[width=0.06\textwidth]{S_min.svg} &  \includesvg[width=0.06\textwidth]{S_max.svg}
  \\
   \SetCell[c=1]{c}{
  \mbox{\small\( \displaystyle 2.02(3)\)} \\
  \mbox{\small\( \displaystyle \{2.0\}\)}} & \SetCell[c=1]{c}{ \vbox{\parshape=1 -0.7cm 3cm {\small{\( \displaystyle \displaystyle \langle \sigma\rangle\langle\partial \tau\rangle- \langle\tau\rangle\langle\partial\sigma\rangle\)}}} \\ \mbox{\small \{3\}} } &
  \includesvg[width=0.06\textwidth]{current_min.svg} &
  \includesvg[width=0.06\textwidth]{current_max.svg} & 
    \SetCell[c=1]{c}{
  \mbox{\small\( \displaystyle 2.20(6)\)} \\
  \mbox{\small\( \displaystyle \{2.222\}\)}} & \SetCell[c=1]{c}{\mbox{\small\(  \displaystyle \langle\partial A\rangle\)} \\ \mbox{\small \{3\}} } & \includesvg[width=0.06\textwidth]{dA_max.svg} &
  \includesvg[width=0.06\textwidth]{dA_min.svg} 
\end{tblr}
    \caption{The Ashkin-Teller Transverse-Field Ising (AT-TFI) and the self-dual Ising-Higgs gauge (SD-IHG) models' three leading operators at their MCPs (multicritical points) identified using the RSMI-NE method, including their scaling dimensions and degeneracies. The expected scaling dimensions of the operators we find are taken from \cite{chester_2019} and \cite{nahum_2021} respectively. The SD-IHG model is conjectured to belong to the $XY^\ast$ universality class and thus to contain a current operator with a scaling dimension of $2.0$ in its conformal spectrum \cite{bonati_2022}. While we find evidence of such an operator for the AT-TFI model (left bottom) we do not find it for the SD-IHG model. All operators show high agreement with their analytical counterparts, both in terms of scaling dimensions and in terms of operator-operator correlations (For more information, see appendix \ref{sec:scaling_dimension}). $\langle\sigma\rangle$ and $\langle\tau\rangle$ denote averaging the $\sigma$ and $\tau$ d.o.fs (see Eq. \ref{eq:H_ATTFI}) over the $3D$ block tempo-spatial block in the AT-TFI case. $\langle A\rangle$ and $\langle S\rangle$ denote averaging the anti-symmetric and symmetric operators $A,S$ (see Eq. \ref{eq:AS}) over the $3D$ spatial block in the SD-IHG case. The derivative operator $\partial$ acts as a lattice discrete derivative on the respective d.o.f (finite-difference). Local configurations that maximize/minimize the neural operators are shown as $2D$ projections (Where the blue and red colors denote the spin states). In the AT-TFI projection, the $\tau$ sub-lattice appears as plaquettes for clarity (the black edges appear as a visual aid and carry no physical meaning). In the SD-IHG case, the projection contains the gauge invariant plaquettes' and bonds' values.  More details regarding the neural operators' projections appear in appendix \ref{sec:projection}.}
    \label{tbl:SDIHG_and_ATTFI_CFT}
\end{table*}

The method has already proven successful in $2D$ systems, including interacting spin and dimer models, on regular and aperiodic lattices \cite{ringel_2018,efe_2021,efe2_2021,efe_2023}. Its advantage is two-fold: first, unlike standard computations of the critical exponents, it provides a complete signature of the underlying universality. Second, unlike exact diagonalization approaches, \emph{e.g.~}the critical torus energy spectrum (CTES) \cite{schuler_2016}, it does not scale exponentially with the system size.

In this work, we refine and extend the RSMI-NE in two ways: in contrast to the previous works we make essential use of non-linear neural operators; we also introduce additional steps in the algorithm providing better numerical control over the information optimization dynamics. A key ingredient in our approach is a successive projection-out of leading operators to target the sub-leading ones. These technical developments and implementation details are described in detail in appendix \ref{sec:RSMINE}.

Data generation for RSMI-NE as well as scaling dimensions analysis were performed using Monte Carlo methods. For the SD-IHG model, we follow \cite{nahum_2021} for an efficient single plaquette and bonds update scheme, without fixing the gauge. For the AT-TFI model, we performed cluster updates, similar to \cite{schuler_2022}. We also employed an additional parallel tempering method to improve the scanning efficiency of the phase space. Scaling dimensions were computed by measuring the Widom scaling of the desired operator's two-point function at the critical point for varying system sizes. For more information, see appendix \ref{sec:scaling_dimension}.

\textit{Results - }We first apply the extended RSMI-NE method to the well-understood case of the AT-TFI model. We extracted the first five leading primary operators (and their degeneracies), including the current operator, \emph{i.e.~}, the operator of scaling dimension $2$. Higher operators can also be constructed but are less relevant in the context of the field-theoretic problem addressed in this paper.

The left column of Table~\ref{tbl:SDIHG_and_ATTFI_CFT} presents the computed scaling dimensions of the primary operators extracted, which are in agreement with the theoretically expected
values of the $U(1)$ theory \cite{chester_2019}. Furthermore, by computing operator-operator long-range correlations between a neural operator and the analytically known operators, we could unambiguously identify the neural operator: Either as some superposition of known operators or as a yet unknown one (For the full identification procedure, see appendix \ref{sec:scaling_dimension}).

Moreover, the scaling operators, parameterized by neural networks, can be accessed directly, rather than just through their exponents. Owing to their non-linearity the analysis is, however, more complicated than in \emph{e.g.~}\cite{efe2_2021}. To understand their action on the local degrees of freedom, we calculate configurations that extremize the values for the numerical operator. These $(2+1)D$ configurations are shown in Table  \ref{tbl:SDIHG_and_ATTFI_CFT} as a $2D$ projection on a spatial plane.  As an example, the $\langle \sigma \rangle^2+\langle \tau \rangle^2$ minimal configuration is that where both $\langle \sigma \rangle$ and $\langle \tau \rangle$ are zero. As such, both the plaquettes and the sites are almost evenly distributed between blue and red (denoting the two possible values of the d.o.f). The maximal configuration is attained where both $\langle \sigma \rangle$ and $\langle \tau \rangle$ are equal to $\pm 1$. As such, the color of the plaquette and the color of the sites is constant.
Indeed, the representative extremal configurations thus obtained can also be seen to extremize the values of the apriori known analytical form of the operator (see caption of Table~\ref{tbl:SDIHG_and_ATTFI_CFT}).

Having validated the method and its ability to detect non-linear relevant operators in the CFT spectrum, particularly the emergent currents, we now apply it to the intriguing case of the SD-IHG model. Here, we can detect the three leading relevant operators in the CFT spectrum, namely the $S$ and $A$ primary operators and the descendants (spatial derivatives) of $A$, with their respective degeneracies. The operators generally act non-linearly on the gauge-invariant constituents, \emph{i.e.~}plaquettes and bonds. 

As before, the right column of Table~\ref{tbl:SDIHG_and_ATTFI_CFT} shows a very good correspondence between the scaling dimensions of the two leading neural operators, and the values of the theoretically expected operators, $A$ and $S$. Strikingly, however, the next three neural operators are inconsistent with a conjectured current operator \cite{bonati_2022}. They are instead the descendants of A and have a higher scaling dimension, whose value of $2.20(6)$ is consistent with that obtained in \cite{nahum_2021}. No operator with the characteristics of a current (namely a vector operator with a scaling dimension of $2$) has been found. 
Its absence in the RSMI-NE results (in contrast to the AT-TFI case) is a strong indication that, in fact, no such operator exists, and, therefore, the self-dual MCP of the SD-IHG theory does not belong to the $XY^*$ universality class.

One might worry that sub-optimal choices of certain hyperparameters inherent to ML methods might prevent the algorithm from finding solutions (operators) that are nevertheless part of the conformal spectrum of the SD-IHG model, particularly the current. However, this is unlikely, as the algorithm identifies operators not only with a lower but also with \emph{higher} scaling dimension than the sought-after current, which, on theoretical grounds, are \emph{harder} to find \cite{gordon_2021}. Further, operators for which the minimal spatial support in terms of macroscopic lattice exceeds the block size $\mathscr{V}$ would also be absent from the computed spectrum, but this can be probed by varying the block and buffer size. We took care to avoid such pitfalls by performing grid optimization of hyperparameters as well as stability to block and buffer size. Furthermore, we carried out a parallel analysis of the AT-TFI model.

\textit{Conclusions and Outlook - }We demonstrated that recently developed numerical RG methods based on information theory and ML can be brought to a level where they shed light on the current open questions in field theory. In particular, we provide strong evidence against the hypothesis that the multicritical point of the $(2+1)D$ self-dual Ising-Higgs gauge theory belongs to the $XY^*$ universality class, showcasing the ability of our extension of the RSMI-NE algorithm to extract conformal data for high-dimensional systems, including sub-leading and descendant operators. We hypothesize that the MCP's universality class can either be of an ``$XY^{**}$" type, where the current operator appears only through higher multiplets, or of a completely novel type.

We expect that the RSMI-NE procedure and its refinements will be a valuable addition to the arsenal of numerical tools in statistical physics and field theory, especially for critical phenomena beyond the Landau paradigm. Apart from extracting the conformal tower including operators that are difficult to resolve by symmetry, it can guide the construction of field-theoretical description by providing microscopic interpretations for the most relevant degrees of freedom.   

\textit{Acknowledgements - }We thank Adam Nahum for fruitful discussions and Doruk Efe Gökmen for his invaluable advice and contribution to the RSMI-NE code. L.O. acknowledges support from the Milner Foundation. S.G. acknowledges support from the Israel Science Foundation (ISF) Grant no. 586/22. Z.R. acknowledges support from ISF Grant 2250/19. M.K.-J. gratefully acknowledges financial support from the European Union’s Horizon 2020 programme under Marie Sklodowska-Curie Grant Agreement No. 896004 (COMPLEX ML).
\bibliography{bibliography}

\begin{thebibliography}{43}%
\makeatletter
\providecommand \@ifxundefined [1]{%
 \@ifx{#1\undefined}
}%
\providecommand \@ifnum [1]{%
 \ifnum #1\expandafter \@firstoftwo
 \else \expandafter \@secondoftwo
 \fi
}%
\providecommand \@ifx [1]{%
 \ifx #1\expandafter \@firstoftwo
 \else \expandafter \@secondoftwo
 \fi
}%
\providecommand \natexlab [1]{#1}%
\providecommand \enquote  [1]{``#1''}%
\providecommand \bibnamefont  [1]{#1}%
\providecommand \bibfnamefont [1]{#1}%
\providecommand \citenamefont [1]{#1}%
\providecommand \href@noop [0]{\@secondoftwo}%
\providecommand \href [0]{\begingroup \@sanitize@url \@href}%
\providecommand \@href[1]{\@@startlink{#1}\@@href}%
\providecommand \@@href[1]{\endgroup#1\@@endlink}%
\providecommand \@sanitize@url [0]{\catcode `\\12\catcode `\$12\catcode
  `\&12\catcode `\#12\catcode `\^12\catcode `\_12\catcode `\%12\relax}%
\providecommand \@@startlink[1]{}%
\providecommand \@@endlink[0]{}%
\providecommand \url  [0]{\begingroup\@sanitize@url \@url }%
\providecommand \@url [1]{\endgroup\@href {#1}{\urlprefix }}%
\providecommand \urlprefix  [0]{URL }%
\providecommand \Eprint [0]{\href }%
\providecommand \doibase [0]{https://doi.org/}%
\providecommand \selectlanguage [0]{\@gobble}%
\providecommand \bibinfo  [0]{\@secondoftwo}%
\providecommand \bibfield  [0]{\@secondoftwo}%
\providecommand \translation [1]{[#1]}%
\providecommand \BibitemOpen [0]{}%
\providecommand \bibitemStop [0]{}%
\providecommand \bibitemNoStop [0]{.\EOS\space}%
\providecommand \EOS [0]{\spacefactor3000\relax}%
\providecommand \BibitemShut  [1]{\csname bibitem#1\endcsname}%
\let\auto@bib@innerbib\@empty
\bibitem [{\citenamefont {Landau}\ and\ \citenamefont
  {Lifshitz}(1980)}]{landau_1980}%
  \BibitemOpen
  \bibfield  {author} {\bibinfo {author} {\bibfnamefont {L.}~\bibnamefont
  {Landau}}\ and\ \bibinfo {author} {\bibfnamefont {E.}~\bibnamefont
  {Lifshitz}},\ }\href@noop {} {\emph {\bibinfo {title} {{Statistical Physics:
  Volume 5 (3rd Edition)}}}}\ (\bibinfo  {publisher} {Butterworth-Heinemann},\
  \bibinfo {year} {1980})\BibitemShut {NoStop}%
\bibitem [{\citenamefont {Sachdev}\ and\ \citenamefont
  {Yin}(2010)}]{sachdev_2010}%
  \BibitemOpen
  \bibfield  {author} {\bibinfo {author} {\bibfnamefont {S.}~\bibnamefont
  {Sachdev}}\ and\ \bibinfo {author} {\bibfnamefont {X.}~\bibnamefont {Yin}},\
  }\href {https://doi.org/10.1016/j.aop.2009.08.003} {\bibfield  {journal}
  {\bibinfo  {journal} {Annals of Physics}\ }\textbf {\bibinfo {volume}
  {325}},\ \bibinfo {pages} {2} (\bibinfo {year} {2010})}\BibitemShut {NoStop}%
\bibitem [{\citenamefont {Xu}(2012)}]{xu_2012}%
  \BibitemOpen
  \bibfield  {author} {\bibinfo {author} {\bibfnamefont {C.}~\bibnamefont
  {Xu}},\ }\href {https://doi.org/10.1142/S0217979212300071} {\bibfield
  {journal} {\bibinfo  {journal} {{International Journal of Modern Physics B}}\
  }\textbf {\bibinfo {volume} {26}},\ \bibinfo {pages} {1230007} (\bibinfo
  {year} {2012})}\BibitemShut {NoStop}%
\bibitem [{\citenamefont {Gang}(2007)}]{xiao_2007}%
  \BibitemOpen
  \bibfield  {author} {\bibinfo {author} {\bibfnamefont {W.~X.}\ \bibnamefont
  {Gang}},\ }\bibinfo {title} {{Quantum field theory of many-body systems: from
  the origin of sound to an origin of light and electrons}}\ (\bibinfo
  {publisher} {Oxford University Press},\ \bibinfo {address} {Oxford},\
  \bibinfo {year} {2007})\ pp.\ \bibinfo {pages} {354--440}\BibitemShut
  {NoStop}%
\bibitem [{\citenamefont {Girvin}(2005)}]{girvin_2005}%
  \BibitemOpen
  \bibfield  {author} {\bibinfo {author} {\bibfnamefont {S.~M.}\ \bibnamefont
  {Girvin}},\ }\bibinfo {title} {{Introduction to the Fractional Quantum Hall
  Effect}},\ in\ \href {https://doi.org/10.1007/3-7643-7393-8_4} {\emph
  {\bibinfo {booktitle} {The Quantum Hall Effect: Poincaré Seminar 2004}}}\
  (\bibinfo  {publisher} {Birkhauser Basel},\ \bibinfo {address} {Basel},\
  \bibinfo {year} {2005})\ pp.\ \bibinfo {pages} {133--162}\BibitemShut
  {NoStop}%
\bibitem [{\citenamefont {Cheng}\ \emph {et~al.}(2016)\citenamefont {Cheng},
  \citenamefont {Zaletel}, \citenamefont {Barkeshli}, \citenamefont
  {Vishwanath},\ and\ \citenamefont {Bonderson}}]{barkeshli_2016}%
  \BibitemOpen
  \bibfield  {author} {\bibinfo {author} {\bibfnamefont {M.}~\bibnamefont
  {Cheng}}, \bibinfo {author} {\bibfnamefont {M.}~\bibnamefont {Zaletel}},
  \bibinfo {author} {\bibfnamefont {M.}~\bibnamefont {Barkeshli}}, \bibinfo
  {author} {\bibfnamefont {A.}~\bibnamefont {Vishwanath}},\ and\ \bibinfo
  {author} {\bibfnamefont {P.}~\bibnamefont {Bonderson}},\ }\href
  {https://doi.org/10.1103/PhysRevX.6.041068} {\bibfield  {journal} {\bibinfo
  {journal} {Phys. Rev. X}\ }\textbf {\bibinfo {volume} {6}},\ \bibinfo {pages}
  {041068} (\bibinfo {year} {2016})}\BibitemShut {NoStop}%
\bibitem [{\citenamefont {Barkeshli}\ \emph {et~al.}(2019)\citenamefont
  {Barkeshli}, \citenamefont {Bonderson}, \citenamefont {Cheng},\ and\
  \citenamefont {Wang}}]{barkeshli_2019}%
  \BibitemOpen
  \bibfield  {author} {\bibinfo {author} {\bibfnamefont {M.}~\bibnamefont
  {Barkeshli}}, \bibinfo {author} {\bibfnamefont {P.}~\bibnamefont
  {Bonderson}}, \bibinfo {author} {\bibfnamefont {M.}~\bibnamefont {Cheng}},\
  and\ \bibinfo {author} {\bibfnamefont {Z.}~\bibnamefont {Wang}},\ }\href
  {https://doi.org/10.1103/PhysRevB.100.115147} {\bibfield  {journal} {\bibinfo
   {journal} {Phys. Rev. B}\ }\textbf {\bibinfo {volume} {100}},\ \bibinfo
  {pages} {115147} (\bibinfo {year} {2019})}\BibitemShut {NoStop}%
\bibitem [{\citenamefont {Senthil}\ \emph {et~al.}(2004)\citenamefont
  {Senthil}, \citenamefont {Balents}, \citenamefont {Sachdev}, \citenamefont
  {Vishwanath},\ and\ \citenamefont {Fisher}}]{senthil_2004}%
  \BibitemOpen
  \bibfield  {author} {\bibinfo {author} {\bibfnamefont {T.}~\bibnamefont
  {Senthil}}, \bibinfo {author} {\bibfnamefont {L.}~\bibnamefont {Balents}},
  \bibinfo {author} {\bibfnamefont {S.}~\bibnamefont {Sachdev}}, \bibinfo
  {author} {\bibfnamefont {A.}~\bibnamefont {Vishwanath}},\ and\ \bibinfo
  {author} {\bibfnamefont {M.~P.~A.}\ \bibnamefont {Fisher}},\ }\bibfield
  {journal} {\bibinfo  {journal} {Physical Review B}\ }\textbf {\bibinfo
  {volume} {70}},\ \href {https://doi.org/10.1103/physrevb.70.144407}
  {10.1103/physrevb.70.144407} (\bibinfo {year} {2004})\BibitemShut {NoStop}%
\bibitem [{\citenamefont {Sachdev}(2018)}]{sachdev_2019}%
  \BibitemOpen
  \bibfield  {author} {\bibinfo {author} {\bibfnamefont {S.}~\bibnamefont
  {Sachdev}},\ }\href {https://doi.org/10.1088/1361-6633/aae110} {\bibfield
  {journal} {\bibinfo  {journal} {Reports on Progress in Physics}\ }\textbf
  {\bibinfo {volume} {82}},\ \bibinfo {pages} {014001} (\bibinfo {year}
  {2018})}\BibitemShut {NoStop}%
\bibitem [{\citenamefont {Sachdev}(2023)}]{sachdev_2023}%
  \BibitemOpen
  \bibfield  {author} {\bibinfo {author} {\bibfnamefont {S.}~\bibnamefont
  {Sachdev}},\ }\bibinfo {title} {{Fractionalization and emergent gauge fields
  I}},\ in\ \href@noop {} {\emph {\bibinfo {booktitle} {Quantum Phases of
  Matter}}}\ (\bibinfo  {publisher} {Cambridge University Press},\ \bibinfo
  {year} {2023})\ p.\ \bibinfo {pages} {151–232}\BibitemShut {NoStop}%
\bibitem [{\citenamefont {Dupont}\ \emph {et~al.}(2021)\citenamefont {Dupont},
  \citenamefont {Gazit},\ and\ \citenamefont {Scaffidi}}]{gazit_2023}%
  \BibitemOpen
  \bibfield  {author} {\bibinfo {author} {\bibfnamefont {M.}~\bibnamefont
  {Dupont}}, \bibinfo {author} {\bibfnamefont {S.}~\bibnamefont {Gazit}},\ and\
  \bibinfo {author} {\bibfnamefont {T.}~\bibnamefont {Scaffidi}},\ }\href
  {https://doi.org/10.1103/PhysRevB.103.L140412} {\bibfield  {journal}
  {\bibinfo  {journal} {Phys. Rev. B}\ }\textbf {\bibinfo {volume} {103}},\
  \bibinfo {pages} {L140412} (\bibinfo {year} {2021})}\BibitemShut {NoStop}%
\bibitem [{\citenamefont {Fradkin}\ and\ \citenamefont
  {Shenker}(1979)}]{fradkin_1979}%
  \BibitemOpen
  \bibfield  {author} {\bibinfo {author} {\bibfnamefont {E.}~\bibnamefont
  {Fradkin}}\ and\ \bibinfo {author} {\bibfnamefont {S.~H.}\ \bibnamefont
  {Shenker}},\ }\href {https://doi.org/10.1103/PhysRevD.19.3682} {\bibfield
  {journal} {\bibinfo  {journal} {Phys. Rev. D}\ }\textbf {\bibinfo {volume}
  {19}},\ \bibinfo {pages} {3682} (\bibinfo {year} {1979})}\BibitemShut
  {NoStop}%
\bibitem [{\citenamefont {Kitaev}(2003)}]{kitaev_2003}%
  \BibitemOpen
  \bibfield  {author} {\bibinfo {author} {\bibfnamefont {A.}~\bibnamefont
  {Kitaev}},\ }\href
  {https://doi.org/https://doi.org/10.1016/S0003-4916(02)00018-0} {\bibfield
  {journal} {\bibinfo  {journal} {Annals of Physics}\ }\textbf {\bibinfo
  {volume} {303}},\ \bibinfo {pages} {2} (\bibinfo {year} {2003})}\BibitemShut
  {NoStop}%
\bibitem [{\citenamefont {Vidal}\ \emph {et~al.}(2009)\citenamefont {Vidal},
  \citenamefont {Dusuel},\ and\ \citenamefont {Schmidt}}]{vidal_2009}%
  \BibitemOpen
  \bibfield  {author} {\bibinfo {author} {\bibfnamefont {J.}~\bibnamefont
  {Vidal}}, \bibinfo {author} {\bibfnamefont {S.}~\bibnamefont {Dusuel}},\ and\
  \bibinfo {author} {\bibfnamefont {K.~P.}\ \bibnamefont {Schmidt}},\
  }\bibfield  {journal} {\bibinfo  {journal} {Physical Review B}\ }\textbf
  {\bibinfo {volume} {79}},\ \href {https://doi.org/10.1103/physrevb.79.033109}
  {10.1103/physrevb.79.033109} (\bibinfo {year} {2009})\BibitemShut {NoStop}%
\bibitem [{\citenamefont {Tupitsyn}\ \emph {et~al.}(2010)\citenamefont
  {Tupitsyn}, \citenamefont {Kitaev}, \citenamefont {Prokof'ev},\ and\
  \citenamefont {Stamp}}]{tupitsyn_2010}%
  \BibitemOpen
  \bibfield  {author} {\bibinfo {author} {\bibfnamefont {I.~S.}\ \bibnamefont
  {Tupitsyn}}, \bibinfo {author} {\bibfnamefont {A.}~\bibnamefont {Kitaev}},
  \bibinfo {author} {\bibfnamefont {N.~V.}\ \bibnamefont {Prokof'ev}},\ and\
  \bibinfo {author} {\bibfnamefont {P.~C.~E.}\ \bibnamefont {Stamp}},\
  }\bibfield  {journal} {\bibinfo  {journal} {Physical Review B}\ }\textbf
  {\bibinfo {volume} {82}},\ \href {https://doi.org/10.1103/physrevb.82.085114}
  {10.1103/physrevb.82.085114} (\bibinfo {year} {2010})\BibitemShut {NoStop}%
\bibitem [{\citenamefont {Gazit}\ \emph {et~al.}(2018)\citenamefont {Gazit},
  \citenamefont {Assaad}, \citenamefont {Sachdev}, \citenamefont {Vishwanath},\
  and\ \citenamefont {Wang}}]{gazit_2018}%
  \BibitemOpen
  \bibfield  {author} {\bibinfo {author} {\bibfnamefont {S.}~\bibnamefont
  {Gazit}}, \bibinfo {author} {\bibfnamefont {F.~F.}\ \bibnamefont {Assaad}},
  \bibinfo {author} {\bibfnamefont {S.}~\bibnamefont {Sachdev}}, \bibinfo
  {author} {\bibfnamefont {A.}~\bibnamefont {Vishwanath}},\ and\ \bibinfo
  {author} {\bibfnamefont {C.}~\bibnamefont {Wang}},\ }\href
  {https://doi.org/10.1073/pnas.1806338115} {\bibfield  {journal} {\bibinfo
  {journal} {Proceedings of the National Academy of Sciences}\ }\textbf
  {\bibinfo {volume} {115}} (\bibinfo {year} {2018})}\BibitemShut {NoStop}%
\bibitem [{\citenamefont {Somoza}\ \emph {et~al.}(2021)\citenamefont {Somoza},
  \citenamefont {Serna},\ and\ \citenamefont {Nahum}}]{nahum_2021}%
  \BibitemOpen
  \bibfield  {author} {\bibinfo {author} {\bibfnamefont {A.~M.}\ \bibnamefont
  {Somoza}}, \bibinfo {author} {\bibfnamefont {P.}~\bibnamefont {Serna}},\ and\
  \bibinfo {author} {\bibfnamefont {A.}~\bibnamefont {Nahum}},\ }\href
  {https://doi.org/10.1103/PhysRevX.11.041008} {\bibfield  {journal} {\bibinfo
  {journal} {Phys. Rev. X}\ }\textbf {\bibinfo {volume} {11}},\ \bibinfo
  {pages} {041008} (\bibinfo {year} {2021})}\BibitemShut {NoStop}%
\bibitem [{\citenamefont {Iqbal}\ and\ \citenamefont
  {McGreevy}(2022)}]{iqbal_2022}%
  \BibitemOpen
  \bibfield  {author} {\bibinfo {author} {\bibfnamefont {N.}~\bibnamefont
  {Iqbal}}\ and\ \bibinfo {author} {\bibfnamefont {J.}~\bibnamefont
  {McGreevy}},\ }\href {https://doi.org/10.21468/SciPostPhys.13.5.114}
  {\bibfield  {journal} {\bibinfo  {journal} {SciPost Phys.}\ }\textbf
  {\bibinfo {volume} {13}},\ \bibinfo {pages} {114} (\bibinfo {year}
  {2022})}\BibitemShut {NoStop}%
\bibitem [{\citenamefont {Bonati}\ \emph
  {et~al.}(2022{\natexlab{a}})\citenamefont {Bonati}, \citenamefont
  {Pelissetto},\ and\ \citenamefont {Vicari}}]{bonati_2022_0}%
  \BibitemOpen
  \bibfield  {author} {\bibinfo {author} {\bibfnamefont {C.}~\bibnamefont
  {Bonati}}, \bibinfo {author} {\bibfnamefont {A.}~\bibnamefont {Pelissetto}},\
  and\ \bibinfo {author} {\bibfnamefont {E.}~\bibnamefont {Vicari}},\ }\href
  {https://doi.org/10.1103/PhysRevE.105.054132} {\bibfield  {journal} {\bibinfo
   {journal} {Phys. Rev. E}\ }\textbf {\bibinfo {volume} {105}},\ \bibinfo
  {pages} {054132} (\bibinfo {year} {2022}{\natexlab{a}})}\BibitemShut
  {NoStop}%
\bibitem [{\citenamefont {Bonati}\ \emph
  {et~al.}(2022{\natexlab{b}})\citenamefont {Bonati}, \citenamefont
  {Pelissetto},\ and\ \citenamefont {Vicari}}]{bonati_2022}%
  \BibitemOpen
  \bibfield  {author} {\bibinfo {author} {\bibfnamefont {C.}~\bibnamefont
  {Bonati}}, \bibinfo {author} {\bibfnamefont {A.}~\bibnamefont {Pelissetto}},\
  and\ \bibinfo {author} {\bibfnamefont {E.}~\bibnamefont {Vicari}},\
  }\bibfield  {journal} {\bibinfo  {journal} {Physical Review B}\ }\textbf
  {\bibinfo {volume} {105}},\ \href
  {https://doi.org/10.1103/physrevb.105.165138} {10.1103/physrevb.105.165138}
  (\bibinfo {year} {2022}{\natexlab{b}})\BibitemShut {NoStop}%
\bibitem [{\citenamefont {Manoj}\ and\ \citenamefont
  {Shenoy}(2023)}]{manoj_2023}%
  \BibitemOpen
  \bibfield  {author} {\bibinfo {author} {\bibfnamefont {N.}~\bibnamefont
  {Manoj}}\ and\ \bibinfo {author} {\bibfnamefont {V.~B.}\ \bibnamefont
  {Shenoy}},\ }\href {https://doi.org/10.1103/PhysRevB.107.165136} {\bibfield
  {journal} {\bibinfo  {journal} {Phys. Rev. B}\ }\textbf {\bibinfo {volume}
  {107}},\ \bibinfo {pages} {165136} (\bibinfo {year} {2023})}\BibitemShut
  {NoStop}%
\bibitem [{\citenamefont {Isakov}\ \emph {et~al.}(2012)\citenamefont {Isakov},
  \citenamefont {Melko},\ and\ \citenamefont {Hastings}}]{isakov_2012}%
  \BibitemOpen
  \bibfield  {author} {\bibinfo {author} {\bibfnamefont {S.~V.}\ \bibnamefont
  {Isakov}}, \bibinfo {author} {\bibfnamefont {R.~G.}\ \bibnamefont {Melko}},\
  and\ \bibinfo {author} {\bibfnamefont {M.~B.}\ \bibnamefont {Hastings}},\
  }\href {https://doi.org/10.1126/science.1212207} {\bibfield  {journal}
  {\bibinfo  {journal} {Science}\ }\textbf {\bibinfo {volume} {335}},\ \bibinfo
  {pages} {193} (\bibinfo {year} {2012})}\BibitemShut {NoStop}%
\bibitem [{\citenamefont {Cardy}(1986)}]{cardy_1986}%
  \BibitemOpen
  \bibfield  {author} {\bibinfo {author} {\bibfnamefont {J.~L.}\ \bibnamefont
  {Cardy}},\ }\href
  {https://doi.org/https://doi.org/10.1016/0550-3213(86)90552-3} {\bibfield
  {journal} {\bibinfo  {journal} {Nuclear Physics B}\ }\textbf {\bibinfo
  {volume} {270}},\ \bibinfo {pages} {186} (\bibinfo {year}
  {1986})}\BibitemShut {NoStop}%
\bibitem [{\citenamefont {Schuler}\ \emph {et~al.}(2023)\citenamefont
  {Schuler}, \citenamefont {Henry}, \citenamefont {Lu},\ and\ \citenamefont
  {Läuchli}}]{schuler_2022}%
  \BibitemOpen
  \bibfield  {author} {\bibinfo {author} {\bibfnamefont {M.}~\bibnamefont
  {Schuler}}, \bibinfo {author} {\bibfnamefont {L.-P.}\ \bibnamefont {Henry}},
  \bibinfo {author} {\bibfnamefont {Y.-M.}\ \bibnamefont {Lu}},\ and\ \bibinfo
  {author} {\bibfnamefont {A.~M.}\ \bibnamefont {Läuchli}},\ }\href
  {https://doi.org/10.21468/SciPostPhys.14.1.001} {\bibfield  {journal}
  {\bibinfo  {journal} {SciPost Phys.}\ }\textbf {\bibinfo {volume} {14}},\
  \bibinfo {pages} {001} (\bibinfo {year} {2023})}\BibitemShut {NoStop}%
\bibitem [{\citenamefont {Zhou}\ \emph {et~al.}(2023)\citenamefont {Zhou},
  \citenamefont {Hu}, \citenamefont {Zhu},\ and\ \citenamefont
  {He}}]{zhou_2023}%
  \BibitemOpen
  \bibfield  {author} {\bibinfo {author} {\bibfnamefont {Z.}~\bibnamefont
  {Zhou}}, \bibinfo {author} {\bibfnamefont {L.}~\bibnamefont {Hu}}, \bibinfo
  {author} {\bibfnamefont {W.}~\bibnamefont {Zhu}},\ and\ \bibinfo {author}
  {\bibfnamefont {Y.-C.}\ \bibnamefont {He}},\ }\href
  {https://doi.org/10.48550/arXiv.2306.16435} {\  (\bibinfo {year}
  {2023})}\BibitemShut {NoStop}%
\bibitem [{\citenamefont {Zhu}\ \emph {et~al.}(2023)\citenamefont {Zhu},
  \citenamefont {Han}, \citenamefont {Huffman}, \citenamefont {Hofmann},\ and\
  \citenamefont {He}}]{zhu_2023}%
  \BibitemOpen
  \bibfield  {author} {\bibinfo {author} {\bibfnamefont {W.}~\bibnamefont
  {Zhu}}, \bibinfo {author} {\bibfnamefont {C.}~\bibnamefont {Han}}, \bibinfo
  {author} {\bibfnamefont {E.}~\bibnamefont {Huffman}}, \bibinfo {author}
  {\bibfnamefont {J.~S.}\ \bibnamefont {Hofmann}},\ and\ \bibinfo {author}
  {\bibfnamefont {Y.-C.}\ \bibnamefont {He}},\ }\bibfield  {journal} {\bibinfo
  {journal} {Physical Review X}\ }\textbf {\bibinfo {volume} {13}},\ \href
  {https://doi.org/10.1103/physrevx.13.021009} {10.1103/physrevx.13.021009}
  (\bibinfo {year} {2023})\BibitemShut {NoStop}%
\bibitem [{\citenamefont {Koch-Janusz}\ and\ \citenamefont
  {Ringel}(2018)}]{ringel_2018}%
  \BibitemOpen
  \bibfield  {author} {\bibinfo {author} {\bibfnamefont {M.}~\bibnamefont
  {Koch-Janusz}}\ and\ \bibinfo {author} {\bibfnamefont {Z.}~\bibnamefont
  {Ringel}},\ }\href {https://doi.org/10.1038/s41567-018-0081-4} {\bibfield
  {journal} {\bibinfo  {journal} {Nature Physics}\ }\textbf {\bibinfo {volume}
  {14}},\ \bibinfo {pages} {578} (\bibinfo {year} {2018})}\BibitemShut
  {NoStop}%
\bibitem [{\citenamefont {Li}\ and\ \citenamefont {Wang}(2018)}]{li_2018}%
  \BibitemOpen
  \bibfield  {author} {\bibinfo {author} {\bibfnamefont {S.-H.}\ \bibnamefont
  {Li}}\ and\ \bibinfo {author} {\bibfnamefont {L.}~\bibnamefont {Wang}},\
  }\href {https://doi.org/10.1103/PhysRevLett.121.260601} {\bibfield  {journal}
  {\bibinfo  {journal} {Phys. Rev. Lett.}\ }\textbf {\bibinfo {volume} {121}},\
  \bibinfo {pages} {260601} (\bibinfo {year} {2018})}\BibitemShut {NoStop}%
\bibitem [{\citenamefont {Di~Sante}\ \emph {et~al.}(2022)\citenamefont
  {Di~Sante}, \citenamefont {Medvidovi\ifmmode~\acute{c}\else \'{c}\fi{}},
  \citenamefont {Toschi}, \citenamefont {Sangiovanni}, \citenamefont
  {Franchini}, \citenamefont {Sengupta},\ and\ \citenamefont
  {Millis}}]{sante_2022}%
  \BibitemOpen
  \bibfield  {author} {\bibinfo {author} {\bibfnamefont {D.}~\bibnamefont
  {Di~Sante}}, \bibinfo {author} {\bibfnamefont {M.}~\bibnamefont
  {Medvidovi\ifmmode~\acute{c}\else \'{c}\fi{}}}, \bibinfo {author}
  {\bibfnamefont {A.}~\bibnamefont {Toschi}}, \bibinfo {author} {\bibfnamefont
  {G.}~\bibnamefont {Sangiovanni}}, \bibinfo {author} {\bibfnamefont
  {C.}~\bibnamefont {Franchini}}, \bibinfo {author} {\bibfnamefont {A.~M.}\
  \bibnamefont {Sengupta}},\ and\ \bibinfo {author} {\bibfnamefont {A.~J.}\
  \bibnamefont {Millis}},\ }\href
  {https://doi.org/10.1103/PhysRevLett.129.136402} {\bibfield  {journal}
  {\bibinfo  {journal} {Phys. Rev. Lett.}\ }\textbf {\bibinfo {volume} {129}},\
  \bibinfo {pages} {136402} (\bibinfo {year} {2022})}\BibitemShut {NoStop}%
\bibitem [{\citenamefont {G\"okmen}\ \emph
  {et~al.}(2021{\natexlab{a}})\citenamefont {G\"okmen}, \citenamefont {Ringel},
  \citenamefont {Huber},\ and\ \citenamefont {Koch-Janusz}}]{efe2_2021}%
  \BibitemOpen
  \bibfield  {author} {\bibinfo {author} {\bibfnamefont {D.~E.}\ \bibnamefont
  {G\"okmen}}, \bibinfo {author} {\bibfnamefont {Z.}~\bibnamefont {Ringel}},
  \bibinfo {author} {\bibfnamefont {S.~D.}\ \bibnamefont {Huber}},\ and\
  \bibinfo {author} {\bibfnamefont {M.}~\bibnamefont {Koch-Janusz}},\ }\href
  {https://doi.org/10.1103/PhysRevLett.127.240603} {\bibfield  {journal}
  {\bibinfo  {journal} {Phys. Rev. Lett.}\ }\textbf {\bibinfo {volume} {127}},\
  \bibinfo {pages} {240603} (\bibinfo {year} {2021}{\natexlab{a}})}\BibitemShut
  {NoStop}%
\bibitem [{\citenamefont {Margalit}\ \emph {et~al.}(2022)\citenamefont
  {Margalit}, \citenamefont {Lesser}, \citenamefont {Pereg-Barnea},\ and\
  \citenamefont {Oreg}}]{margalit_2022}%
  \BibitemOpen
  \bibfield  {author} {\bibinfo {author} {\bibfnamefont {G.}~\bibnamefont
  {Margalit}}, \bibinfo {author} {\bibfnamefont {O.}~\bibnamefont {Lesser}},
  \bibinfo {author} {\bibfnamefont {T.}~\bibnamefont {Pereg-Barnea}},\ and\
  \bibinfo {author} {\bibfnamefont {Y.}~\bibnamefont {Oreg}},\ }\href
  {https://doi.org/10.1103/PhysRevB.105.205139} {\bibfield  {journal} {\bibinfo
   {journal} {Phys. Rev. B}\ }\textbf {\bibinfo {volume} {105}},\ \bibinfo
  {pages} {205139} (\bibinfo {year} {2022})}\BibitemShut {NoStop}%
\bibitem [{\citenamefont {Zhang}\ and\ \citenamefont {You}(2023)}]{zhang_2023}%
  \BibitemOpen
  \bibfield  {author} {\bibinfo {author} {\bibfnamefont {Z.}~\bibnamefont
  {Zhang}}\ and\ \bibinfo {author} {\bibfnamefont {Y.-Z.}\ \bibnamefont
  {You}},\ }\href {https://doi.org/10.48550/arXiv.2306.14838} {\bibfield
  {journal} {\bibinfo  {journal} {arXiv preprint arXiv:2306.14838}\ } (\bibinfo
  {year} {2023})}\BibitemShut {NoStop}%
\bibitem [{\citenamefont {Gordon}\ \emph {et~al.}(2021)\citenamefont {Gordon},
  \citenamefont {Banerjee}, \citenamefont {Koch-Janusz},\ and\ \citenamefont
  {Ringel}}]{gordon_2021}%
  \BibitemOpen
  \bibfield  {author} {\bibinfo {author} {\bibfnamefont {A.}~\bibnamefont
  {Gordon}}, \bibinfo {author} {\bibfnamefont {A.}~\bibnamefont {Banerjee}},
  \bibinfo {author} {\bibfnamefont {M.}~\bibnamefont {Koch-Janusz}},\ and\
  \bibinfo {author} {\bibfnamefont {Z.}~\bibnamefont {Ringel}},\ }\bibfield
  {journal} {\bibinfo  {journal} {Physical Review Letters}\ }\textbf {\bibinfo
  {volume} {126}},\ \href {https://doi.org/10.1103/physrevlett.126.240601}
  {10.1103/physrevlett.126.240601} (\bibinfo {year} {2021})\BibitemShut
  {NoStop}%
\bibitem [{\citenamefont {G\"okmen}\ \emph
  {et~al.}(2021{\natexlab{b}})\citenamefont {G\"okmen}, \citenamefont {Ringel},
  \citenamefont {Huber},\ and\ \citenamefont {Koch-Janusz}}]{efe_2021}%
  \BibitemOpen
  \bibfield  {author} {\bibinfo {author} {\bibfnamefont {D.~E.}\ \bibnamefont
  {G\"okmen}}, \bibinfo {author} {\bibfnamefont {Z.}~\bibnamefont {Ringel}},
  \bibinfo {author} {\bibfnamefont {S.~D.}\ \bibnamefont {Huber}},\ and\
  \bibinfo {author} {\bibfnamefont {M.}~\bibnamefont {Koch-Janusz}},\ }\href
  {https://doi.org/10.1103/PhysRevE.104.064106} {\bibfield  {journal} {\bibinfo
   {journal} {Phys. Rev. E}\ }\textbf {\bibinfo {volume} {104}},\ \bibinfo
  {pages} {064106} (\bibinfo {year} {2021}{\natexlab{b}})}\BibitemShut
  {NoStop}%
\bibitem [{\citenamefont {Gökmen}\ \emph {et~al.}(2023)\citenamefont
  {Gökmen}, \citenamefont {Biswas}, \citenamefont {Huber}, \citenamefont
  {Ringel}, \citenamefont {Flicker},\ and\ \citenamefont
  {Koch-Janusz}}]{efe_2023}%
  \BibitemOpen
  \bibfield  {author} {\bibinfo {author} {\bibfnamefont {D.~E.}\ \bibnamefont
  {Gökmen}}, \bibinfo {author} {\bibfnamefont {S.}~\bibnamefont {Biswas}},
  \bibinfo {author} {\bibfnamefont {S.~D.}\ \bibnamefont {Huber}}, \bibinfo
  {author} {\bibfnamefont {Z.}~\bibnamefont {Ringel}}, \bibinfo {author}
  {\bibfnamefont {F.}~\bibnamefont {Flicker}},\ and\ \bibinfo {author}
  {\bibfnamefont {M.}~\bibnamefont {Koch-Janusz}},\ }\href
  {https://doi.org/10.48550/arXiv.2301.11934} {\bibinfo {title} {{Compression
  theory for inhomogeneous systems}}} (\bibinfo {year} {2023})\BibitemShut
  {NoStop}%
\bibitem [{\citenamefont {Tishby}\ \emph {et~al.}(1999)\citenamefont {Tishby},
  \citenamefont {Pereira},\ and\ \citenamefont {Bialek}}]{tishbi_1999}%
  \BibitemOpen
  \bibfield  {author} {\bibinfo {author} {\bibfnamefont {N.}~\bibnamefont
  {Tishby}}, \bibinfo {author} {\bibfnamefont {F.~C.}\ \bibnamefont
  {Pereira}},\ and\ \bibinfo {author} {\bibfnamefont {W.}~\bibnamefont
  {Bialek}},\ }in\ \href {https://arxiv.org/abs/physics/0004057} {\emph
  {\bibinfo {booktitle} {Proc. of the 37-th Annual Allerton Conference on
  Communication, Control and Computing}}}\ (\bibinfo {year} {1999})\ pp.\
  \bibinfo {pages} {368--377}\BibitemShut {NoStop}%
\bibitem [{\citenamefont {van~den Oord}\ \emph {et~al.}(2019)\citenamefont
  {van~den Oord}, \citenamefont {Li},\ and\ \citenamefont
  {Vinyals}}]{oord_2019}%
  \BibitemOpen
  \bibfield  {author} {\bibinfo {author} {\bibfnamefont {A.}~\bibnamefont
  {van~den Oord}}, \bibinfo {author} {\bibfnamefont {Y.}~\bibnamefont {Li}},\
  and\ \bibinfo {author} {\bibfnamefont {O.}~\bibnamefont {Vinyals}},\ }\href
  {https://doi.org/10.48550/arXiv.1807.03748} {\bibinfo {title}
  {{Representation Learning with Contrastive Predictive Coding}}} (\bibinfo
  {year} {2019})\BibitemShut {NoStop}%
\bibitem [{\citenamefont {Belghazi}\ \emph {et~al.}(2018)\citenamefont
  {Belghazi}, \citenamefont {Baratin}, \citenamefont {Rajeshwar}, \citenamefont
  {Ozair}, \citenamefont {Bengio}, \citenamefont {Courville},\ and\
  \citenamefont {Hjelm}}]{mine_2018}%
  \BibitemOpen
  \bibfield  {author} {\bibinfo {author} {\bibfnamefont {M.~I.}\ \bibnamefont
  {Belghazi}}, \bibinfo {author} {\bibfnamefont {A.}~\bibnamefont {Baratin}},
  \bibinfo {author} {\bibfnamefont {S.}~\bibnamefont {Rajeshwar}}, \bibinfo
  {author} {\bibfnamefont {S.}~\bibnamefont {Ozair}}, \bibinfo {author}
  {\bibfnamefont {Y.}~\bibnamefont {Bengio}}, \bibinfo {author} {\bibfnamefont
  {A.}~\bibnamefont {Courville}},\ and\ \bibinfo {author} {\bibfnamefont
  {D.}~\bibnamefont {Hjelm}},\ }in\ \href
  {https://proceedings.mlr.press/v80/belghazi18a.html} {\emph {\bibinfo
  {booktitle} {Proceedings of the 35th International Conference on Machine
  Learning}}},\ \bibinfo {series} {Proceedings of Machine Learning Research},
  Vol.~\bibinfo {volume} {80},\ \bibinfo {editor} {edited by\ \bibinfo {editor}
  {\bibfnamefont {J.}~\bibnamefont {Dy}}\ and\ \bibinfo {editor} {\bibfnamefont
  {A.}~\bibnamefont {Krause}}}\ (\bibinfo  {publisher} {PMLR},\ \bibinfo {year}
  {2018})\ pp.\ \bibinfo {pages} {531--540}\BibitemShut {NoStop}%
\bibitem [{\citenamefont {Chester}\ \emph {et~al.}(2020)\citenamefont
  {Chester}, \citenamefont {Landry}, \citenamefont {Liu}, \citenamefont
  {Poland}, \citenamefont {Simmons-Duffin}, \citenamefont {Su},\ and\
  \citenamefont {Vichi}}]{chester_2019}%
  \BibitemOpen
  \bibfield  {author} {\bibinfo {author} {\bibfnamefont {S.~M.}\ \bibnamefont
  {Chester}}, \bibinfo {author} {\bibfnamefont {W.}~\bibnamefont {Landry}},
  \bibinfo {author} {\bibfnamefont {J.}~\bibnamefont {Liu}}, \bibinfo {author}
  {\bibfnamefont {D.}~\bibnamefont {Poland}}, \bibinfo {author} {\bibfnamefont
  {D.}~\bibnamefont {Simmons-Duffin}}, \bibinfo {author} {\bibfnamefont
  {N.}~\bibnamefont {Su}},\ and\ \bibinfo {author} {\bibfnamefont
  {A.}~\bibnamefont {Vichi}},\ }\href {https://doi.org/10.1007/JHEP06(2020)142}
  {\bibfield  {journal} {\bibinfo  {journal} {JHEP}\ }\textbf {\bibinfo
  {volume} {06}},\ \bibinfo {pages} {142}},\ \Eprint
  {https://arxiv.org/abs/1912.03324} {1912.03324} \BibitemShut {NoStop}%
\bibitem [{\citenamefont {Schuler}\ \emph {et~al.}(2016)\citenamefont
  {Schuler}, \citenamefont {Whitsitt}, \citenamefont {Henry}, \citenamefont
  {Sachdev},\ and\ \citenamefont {Läuchli}}]{schuler_2016}%
  \BibitemOpen
  \bibfield  {author} {\bibinfo {author} {\bibfnamefont {M.}~\bibnamefont
  {Schuler}}, \bibinfo {author} {\bibfnamefont {S.}~\bibnamefont {Whitsitt}},
  \bibinfo {author} {\bibfnamefont {L.-P.}\ \bibnamefont {Henry}}, \bibinfo
  {author} {\bibfnamefont {S.}~\bibnamefont {Sachdev}},\ and\ \bibinfo {author}
  {\bibfnamefont {A.~M.}\ \bibnamefont {Läuchli}},\ }\bibfield  {journal}
  {\bibinfo  {journal} {Physical Review Letters}\ }\textbf {\bibinfo {volume}
  {117}},\ \href {https://doi.org/10.1103/physrevlett.117.210401}
  {10.1103/physrevlett.117.210401} (\bibinfo {year} {2016})\BibitemShut
  {NoStop}%
\bibitem [{\citenamefont {Sandvik}\ and\ \citenamefont
  {Zhao}(2020)}]{sandvik_2020}%
  \BibitemOpen
  \bibfield  {author} {\bibinfo {author} {\bibfnamefont {A.~W.}\ \bibnamefont
  {Sandvik}}\ and\ \bibinfo {author} {\bibfnamefont {B.}~\bibnamefont {Zhao}},\
  }\href {https://doi.org/10.1088/0256-307X/37/5/057502} {\bibfield  {journal}
  {\bibinfo  {journal} {Chinese Physics Letters}\ }\textbf {\bibinfo {volume}
  {37}},\ \bibinfo {eid} {057502} (\bibinfo {year} {2020})}\BibitemShut
  {NoStop}%
\bibitem [{\citenamefont {Oppenheim}\ and\ \citenamefont
  {Ringel}(2024)}]{ringel_TBP}%
  \BibitemOpen
  \bibfield  {author} {\bibinfo {author} {\bibfnamefont {L.}~\bibnamefont
  {Oppenheim}}\ and\ \bibinfo {author} {\bibfnamefont {Z.}~\bibnamefont
  {Ringel}},\ }\href@noop {} {\  (\bibinfo {year} {2024})}\BibitemShut
  {NoStop}%
\bibitem [{\citenamefont {Maddison}\ \emph {et~al.}(2016)\citenamefont
  {Maddison}, \citenamefont {Mnih},\ and\ \citenamefont {Teh}}]{maddison_2016}%
  \BibitemOpen
  \bibfield  {author} {\bibinfo {author} {\bibfnamefont {C.~J.}\ \bibnamefont
  {Maddison}}, \bibinfo {author} {\bibfnamefont {A.}~\bibnamefont {Mnih}},\
  and\ \bibinfo {author} {\bibfnamefont {Y.~W.}\ \bibnamefont {Teh}},\ }\href
  {https://doi.org/10.48550/ARXIV.1611.00712} {\bibinfo {title} {{The Concrete
  Distribution: A Continuous Relaxation of Discrete Random Variables}}}
  (\bibinfo {year} {2016})\BibitemShut {NoStop}%
\end{thebibliography}%
\bibliographystyle{apsrev4-2} 
\clearpage

\section{Appendix A - Monte Carlo}
\renewcommand{\theequation}{S.\arabic{equation}}
\setcounter{equation}{0}
\subsection{SD-IHG}
Generating Monte Carlo snapshots of the SD-IHG model for the RSMI-NE algorithm was conducted on a system with linear size $L=18$ (overall $4L^3=2^5 3^6$ d.o.fs) at the MCP ($\mathrm{tanh}(K_c)=0.6367$, $\mathrm{tanh}(J_c)=\frac{1-tanh(K_c)}{1+tanh(K_c}$). After every $10^4$ single-spin sweeps, $18$ pairs of $(\mathscr{V},\mathscr{E})$ were extracted from the Monte Carlo snapshot with varying $\mathscr{V}$ and buffer sizes. Overall, approximately $\sim 10^7$ samples were extracted as a corpus for the RSMI-NE algorithm.

\subsection{AT-TFI}\label{sec:monte_carlo_ATTFI}
\begin{figure}[htp]
    \centering
    \adjustbox{trim={0.0\width} {0.45\height} {0.6\width} {0.0\height},clip}{
    \includegraphics[width=1.2\textwidth]{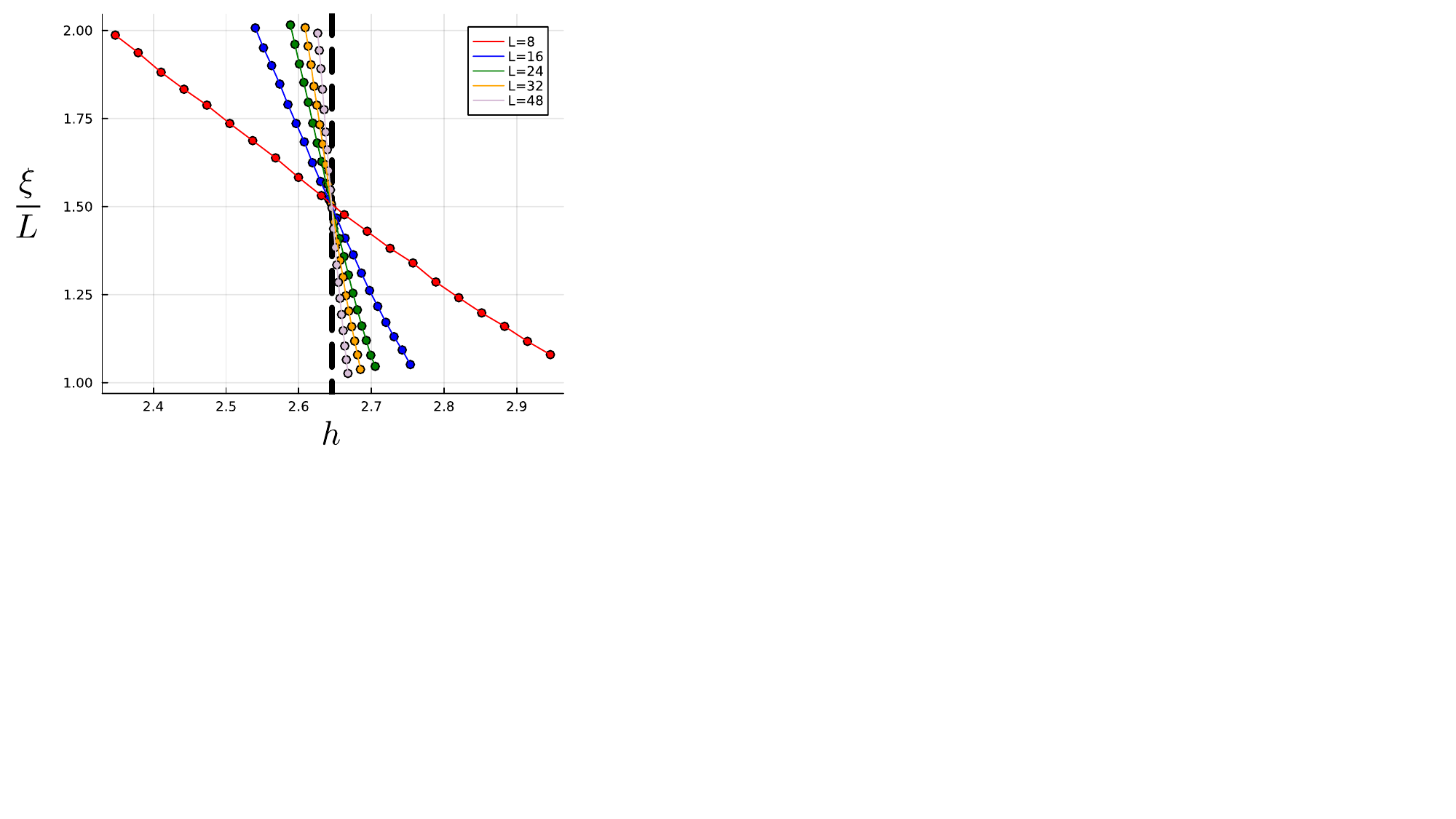}}
    \caption{Finite size scaling of the correlation length for various linear sizes. By collapsing the curves, the critical disordering field is found to be $h_c=2.6458\pm0.0001$ (denoted by a vertical black dashed line).}
    \label{fig:correlation_length_ATTFI}
\end{figure}

Simulation of the AT-TFI model was done using a standard quantum to classical mapping, with effective couplings $\tilde{h}=\frac{1}{2}\log\coth(h\Delta\tau), \tilde{J}=J\Delta\tau,\tilde{J}_{AT}=J_{AT}\Delta \tau$, with a Trotter decomposition step of $\Delta\tau=0.083$ and a imaginary-time axis of length $L_z=3L$ (effective inverse temperature $\beta=L_z\cdot \Delta\tau$). The value for $\Delta\tau$ was chosen such that the ratio between space correlations and time correlation is $3$, which would allow us to scale the time-like axis of $\mathscr{V},\mathscr{E}$ accordingly by an integer factor. For such a choice of parameters, with $J=1, J_{AT}=0.625$, the critical disordering field was calculated to be $h_c=2.6458\pm0.0001$ (See Fig. \ref{fig:correlation_length_ATTFI}).

Generating Monte Carlo samples for the RSMI-NE algorithm was conducted on a system with linear size $L=32$  (overall $L^3=2^{15}$ d.o.fs) at the critical point. After every $10^4$ cluster updates, $32$ pairs of $(\mathscr{V},\mathscr{E})$ were extracted with varying $\mathscr{V}$ and buffer sizes. Overall, approximately $\sim 10^7$ samples were extracted as a corpus for the RSMI-NE algorithm.

\subsection{Identifying extracted operators} \label{sec:scaling_dimension}

For the estimation of the scaling dimension, we have used systems of linear sizes $L=4,6,8,12,16,24$ for the SD-IHG model, and $L=8,10,12,16,24,32$ for the AT-TFI model. At criticality, using the Widom scaling form, we can compute the scaling dimension $\Delta_\mathcal{O}$ for a given operator $\mathcal{O}$  as a log-ratio between the operator antipodal correlation of different system sizes. Given two systems $S_1,S_2$ with corresponding linear sizes $L,2L$ the estimated scaling dimension is given by:
\begin{equation}
    \Delta_\mathcal{O}(L)=\frac{1}{2}\log_{2}\frac{\left\langle \mathcal{O}(0,0,0)\mathcal{O}(\frac{L}{2},\frac{L}{2},\frac{L}{2})\right\rangle _{S_{1}}}{\left\langle \mathcal{O}(0,0,0)\mathcal{O}(\frac{2L}{2},\frac{2L}{2},\frac{2L}{2})\right\rangle _{S_{2}}}
\end{equation}

\begin{figure}[htp]
\centering
\adjustbox{trim={0.00\width} {.45\height} {0.05\width} {0\height},clip}{
\includegraphics[width=0.50\textwidth]{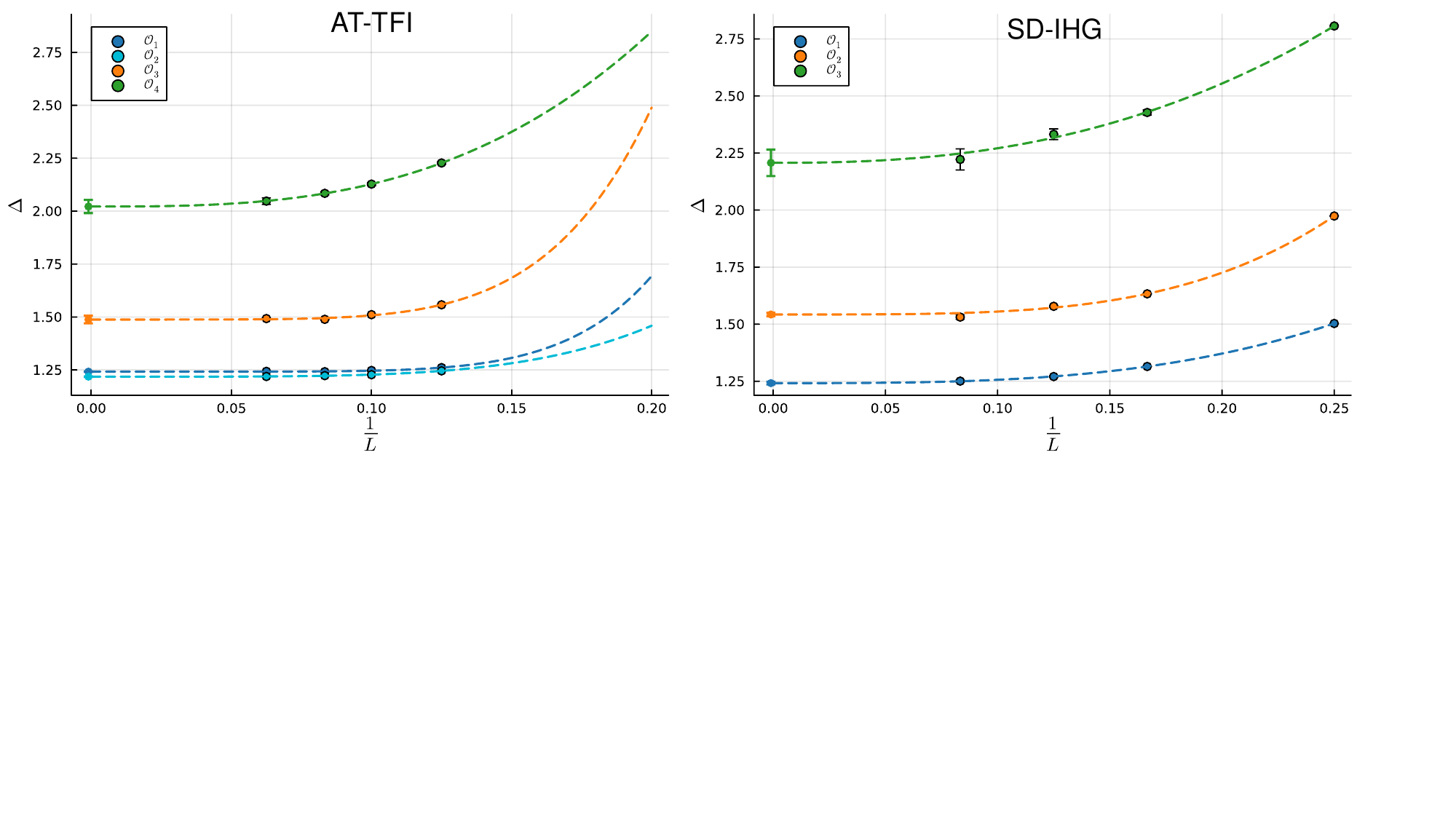}}
\caption{Scaling dimensions for the AT-TFI and SD-IHG models as a function of inverse system size $\frac{1}{L}$. Error bars denote one standard deviation. The crossing of the fitted function (Eq. \ref{eq:widom}) with the vertical axis is the extrapolated scaling dimension at $L\to\infty$.}
    \label{fig:widom}
\end{figure}

The scaling dimensions for the various system sizes were fitted (see Fig. \ref{fig:widom}) by the function:
\begin{equation}
  \Delta_\mathcal{O}(\frac{1}{L})=A(\frac{1}{L})^{\omega}+\Delta_\mathcal{O}(L\to\infty)  
  \label{eq:widom}
\end{equation}
to get the asymptotic scaling dimension of the operator for an infinite system \cite{sandvik_2020}.

Calculating correlation between operators $\mathcal{O}_1$ and $\mathcal{O}_2$  was performed via long-range (Pearson-like) operator-operator correlation:
\begin{gather}
\rho_{\left(\mathcal{O}_1,\mathcal{O}_2\right)}= \\
\frac{\left|\left<\mathcal{O}_1(0,0,0)\mathcal{O}_2(\frac{L}{2},\frac{L}{2},\frac{L}{2})\right>\right|}{\sqrt{\left|\left<\mathcal{O}_1(0,0,0)\mathcal{O}_1(\frac{L}{2},\frac{L}{2},\frac{L}{2})\right>\left<\mathcal{O}_2(0,0,0)\mathcal{O}_2(\frac{L}{2},\frac{L}{2},\frac{L}{2})\right>\right|}}  \nonumber    
\end{gather}
where $0\le\rho_{\left(\mathcal{O}_1,\mathcal{O}_2\right)}\le 1$ measures the similarity between $\mathcal{O}_1$ and $\mathcal{O}_2$. Because of the operator product expansion property of CFTs, we know that for any operator $\mathcal{A}$:
\begin{equation}
\sum_i \rho_{\left(\mathcal{A},\mathcal{O}_i\right)}^2=1    
\end{equation}
where $\mathcal{O}_i$ is the set of all operators in the spectrum. This allows for a complete identification of a given neural operator in terms of the primary and descendant CFT operators.

In general, we find that the procedure produces a highly pristine leading operator within each symmetry sector, as also demonstrated in Fig. \ref{fig:widom}. Nonetheless, we find both experimentally (and theoretically \cite{ringel_TBP} by generalizing the results of \cite{gordon_2021}) that the extracted subleading operators may mix in some small amounts of leading operators within the same symmetry sector. While for the AT-TFI we found this mixing to be negligible, for SD-IHG it had some effect. To circumvent this, we capitalize on the limited form of this mixing and the pristine form of the leading neural operator and project out the leading component. We do so by subtracting the leading neural operator from the sub-leading neural operator such that the long-range correlation with the leading neural operator is reduced to zero. We note by passing that this extra procedure was not required for the current operator as it is leading in its symmetry sector.

\section{Appendix B - RSMI-NE}\label{sec:RSMINE}
\subsection{Description of the basic algorithm and the extensions introduced}
\begin{figure}[htp]
\centering
\adjustbox{trim={0.00\width} {.28\height} {0.3\width} {0\height},clip}{
\includegraphics[width=0.62\textwidth]{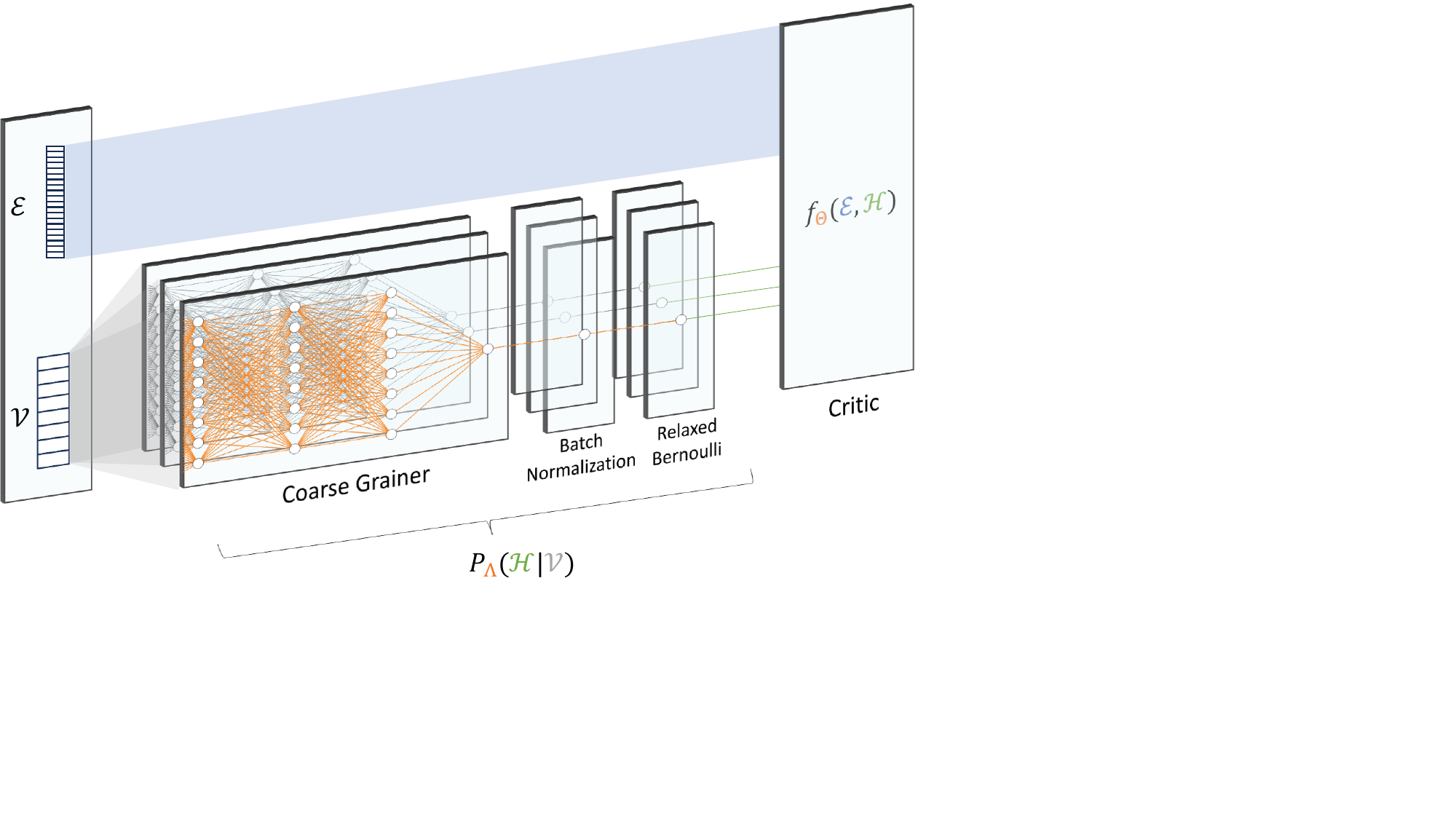}}
\caption{RSMI-NE neural network used to study leading operators. At the first stage, the block $\mathscr{V}$ passes through multiple copies of coarse grainers, batch normalization layers, and Relaxed Bernoulli layers \cite{maddison_2016}, encoding $\mathscr{V}$ into noisy bits $\mathscr{H}$ ($P_\Lambda\left(\mathscr{H}|\mathscr{V}\right)$. The front-most copy shows the operator that is being learned at the current iteration. The two copies behind it show frozen operators from previous training iterations. Correspondingly, the orange edges denote trainable weights, while gray weights denote untrainable weights. In the second stage, all the outputs of the sets $\mathscr{H}$ are fed, together with the environment $\mathscr{E}$, to the INFO-NCE critic $f_\Theta\left(\mathscr{H},\mathscr{E}\right)$ (parameterized by a set of trainable weights $\Theta$).}
    \label{fig:rsmi_visualization}
\end{figure}

The RSMI-NE algorithm used and extended here is based on Refs. \cite{efe_2021,efe2_2021}. The RSMI-NE neural network is composed of two components: A coarse-grainer and a critic. The role of the coarse-grainer, in our case a general fully-connected network parameterized by a set of weights $\Lambda$, is to compress $\mathscr{V}$ to a discrete representation $P_\Lambda(\mathscr{H}|\mathscr{V})$, and the role of the critic (parameterized by a set of weights $\Theta$) is to estimate the mutual information between random variables (which is done via the INFO-NCE method \cite{oord_2019}). The concatenation of the two components (see Fig. \ref{fig:rsmi_visualization}) yields a network that gets as inputs batches of pairs $(\mathscr{V},\mathscr{E})$ sampled from the joint distribution $P(\mathscr{V},\mathscr{E})$, usually by using Monte Carlo sampling, and outputs the mutual information $I_\Theta(\mathscr{H},\mathscr{E})$. The coarse-grainer and the critic are trained together to increase the aforementioned mutual information, which in turn optimizes the coarse-grainer to extract the optimal $\mathscr{H}$. 

The discretization of $\mathscr{H}$ is done via a discretization layer (``Relaxed Bernoulli"), added at the end of the coarse-grainer network, which is analogous to the softmax technique \cite{maddison_2016}. 

In this paper, two major updates were made to the RSMI-NE's architecture and protocol, in order to account for non-linear and sub-leading operators. The success of the RSMI-NE method relies heavily on a good correspondence between the neural operators and the analytical operators in the CFT spectrum. However, as was shown in \cite{efe_2021}, the neural operators will generally consist of some mixture of the analytical operators. There, a principal component analysis (PCA) was performed on an ensemble of linear neural operators, and the leading analytical operators were identified with the leading principal components. Such an analysis becomes an unviable option in the case of non-linear neural operators. Moreover, the use of PCA lacks firm theoretical grounds that will justify this correspondence in a general setting.

Recent theoretical results have shown that a version of RSMI, which limits information not by discretization but by keeping the mapping $P\left(\mathscr{H}|\mathscr{V}\right)$ noisy, corresponds to extracting the pristine leading operator \cite{gordon_2021}. Concretely, such a version of RSMI-NE minimizes the Information-bottleneck (IB) Lagrangian under a fixed cardinality of the representation $\left| \mathscr{H} \right|$  \cite{tishbi_1999}:
\begin{equation}\label{eq_IB}
    \min_{P(\mathscr{H}|\mathscr{V})} \mathcal{L}_{IB}=\min_{P(\mathscr{H}|\mathscr{V})} 
 I\left(\mathscr{V};\mathscr{H}\right)-\beta I\left(\mathscr{H};\mathscr{E}\right),
    \end{equation}
where $\beta$ is a finite parameter that determines the noise level of the mapping. At $\beta\to 0$ the functional is minimized when the mapping retains no information from $\mathscr{V}$. At $\beta\to\infty$ the mapping is noiseless, due to the data processing inequality. In the latter case, minimizing the IB Lagrangian is equivalent to the RSMI-NE method as presented in \cite{efe_2021,efe2_2021}. 

At some critical value $\beta_c>0$ (and for $\left| \mathscr{H}\right| \ge 2$) the IB optimization goes through a bifurcation point, and the solution becomes non-trivial (i.e. the solution becomes $\mathscr{V}$ dependent). As was shown in \cite{gordon_2021}, the solution infinitesimally above $\beta_c$ is analytically related to the leading operator in the CFT spectrum. Thus, by forcing the encoder to learn the leading infinitesimal amount of information on $\mathscr{V}$, pristine operators can be extracted.  

Here we use tunable noise to control the amount of information being learnt. This role is played in IB by the compression penalty term (last term in Eq. \ref{eq_IB}) which however makes numerical optimization of Eq. \ref{eq_IB} difficult in practice. In this paper, the level of compression is tuned in a computationally efficient way by including a batch normalization layer before the discretization layer which controls the variance of ${\mathscr{H}}$. The batch normalization layer takes every mini-batch $B$ (samples of $P\left(\mathscr{H}|\mathscr{V}\right)$ before the discretization), and normalizes it to have a mean $\beta$ and a standard deviation $\gamma$, both of which are trainable parameters:
\begin{gather}
    \mu_B=\frac{1}{\left| B\right|}\sum_{x\in B}x,\sigma_B^2=\frac{1}{\left| B\right|}\sum_{x\in B}(x-\mu_B)^2 \\
    BatchNorm(x;\beta,\gamma)=\gamma\frac{x-\mu_B}{\sqrt{\sigma_B^2+\epsilon}}+\beta \nonumber
\end{gather}
with $\epsilon$ being some small untrainable constant added to maintain numerical stability. Imposing an upper bound on the value of $\gamma$ limits the amount of mutual information $I\left(\mathscr{V};\mathscr{H}\right)$. Such a procedure adds noise to the output bits of the coarse-grainer (``noisy bits") and thus ensures that in accordance with \cite{gordon_2021}, the coarse-grainer is pressured by the optimization process to learn the pristine leading operator.

In order to probe sub-leading operators, we devised two methods: consecutive learning of noisy bits and symmetry projection.
In the first method, the noisy bits in $\mathscr{H}$ are learned in a consecutive manner, where in each step a new noisy bit is learned based on all of the previous ones (``background noisy bits") $P(\mathscr{H}_N|\mathscr{H}_1\mathscr{H}_2\dots\mathscr{H}_{N-1} \mathscr{V})$, which are held constant during the training process of the $N$'th noisy bit. A new critic network is initialized at the beginning of each step. The RSMI-NE neural network is then forced to learn in each step only the operator that yields the biggest change in the overall mutual information, given the background noisy bits. While consecutive learning is inspired by \cite{gordon_2021}, a mathematical proof supporting this would appear in a separate publication \cite{ringel_TBP}. 

The second method makes use of symmetries in order to focus on operators lying in a particular symmetry sector. Given a symmetry group $G$, we can partially symmetrize the dataset by acting on samples of $\mathscr{V}$ (but not of $\mathscr{E}$ with a random element in $G$, which is a form of data augmentation. This symmetrization washes out the information gained from non-symmetric operators in the spectrum. By employing this method, one can directly target the leading operator in the $G$-invariant symmetry sector.

Furthermore, it is also possible to employ the symmetry in order to learn the leading operator which is not invariant under the symmetry. This is done first by augmenting $\mathscr{V}$ and learning all the invariant operators, without adding noise to the bits, until we exhaust all the relevant information in them. Then, we remove the augmentation and learn directly the desired operator. Essentially, by  ``projecting out" the desired operator in the first stage, we allow the RSMI-NE algorithm to learn it in a clear manner in the second stage.

The exact training protocols for the AT-TFI and the SD-IHG models that employ these two methods appear in \ref{sec:application_ATTFI} and \ref{sec:application_SDIHG}. In both cases, the symmetries of the models were used via the symmetry projection method. This allowed us to improve the efficiency of the procedure, reaching the hypothesized (or known) current operator within several steps. However, the symmetry projection is not a mandatory part of the protocol, as sub-leading operators were also found (though less pristine) via a systematic usage of the consecutive learning method alone.

A single stochastic gradient descent (SGD) training step can be summed up in the following pseudo-code block (see also Fig. \ref{fig:rsmi_visualization} for a visual aid):
\begin{tcolorbox}
\begin{tiny}
\begin{verbatim}
Arguments:
(V,E) - Samples from the Monte Carlo snapshots.
        V is a local patch of the system, 
        E is a spatially separated environment of V.
(G) - The group under which the trained neural operator will be invariant.
(tau) - The temperature for the current step. Decays exponentially between steps.
(background noisy bits) - A set of previously trained operators,
                          with their weights frozen (untrainable).

function train_step(V,E,G,tau,background noisy bits)
    choose a random element g in G
    V ← act with g on V
    H ← feed V into the coarse grainer
    H ← feed H into the batch normalization layer
    H ← concatenate H and background noisy bits
    H ← discretize H by a Relaxed Bernoulli layer with temperature tau
                                            (see [43])
    MI ← feed H,E to a neural mutual information estimator (the critic)
                                            (see [37])
    
    perform gradient descent on the parameters of the coarse grainer,
                                                      batch normalization layer 
                                                      and critic 
\end{verbatim}
\end{tiny}
\end{tcolorbox}

\subsection{Hyperparameters for the RSMI-NE}
The geometry for RSMI-NE was chosen as follows: For the AT-TFI model, $\mathscr{V}$ was a block of spins of size $4\times4\times10$ (ratio between space correlations and time correlation is $3$) and the buffer size was taken to be $4\times4\times12$, such that $\mathscr{E}$ consisted of the boundary of a block of spins of size $12 \times 12 \times 34$. For the SD-IHG model, $\mathscr{V}$ was a block containing the gauge invariant bonds ($\tau\sigma\tau$) and the plaquettes ($\sigma\sigma\sigma\sigma$) that are encapsulated by a block of size $2\times 2\times 2$ and the buffer size was taken to be $6 \times 6 \times 6$ such that $\mathscr{E}$ consisted of gauge invariant bonds and plaquettes that are encapsulated by the boundary of a block of size $14 \times 14 \times 14$. Generally, increasing the buffer between $\mathscr{V}$ and $\mathscr{E}$ improves the result (as sub-leading operators are diminished), but takes more training resources (more training steps and larger batch size), as the mutual information between the two variables decreases.

The batch normalization scale was set to a maximal standard deviation of $\gamma_{max}=1.0$. This low value is important to get pristine operators in a symmetry sector that includes multiple symmetry-invariant operators. The batch size was set to 8000.

 The coarse-grainer had three fully connected layers and a width of eight times the input size. The critic, which takes $\mathscr{E},\mathscr{H}$ and outputs their mutual information, includes three stages. In the first stage, $\mathscr{E}$ (a rectangular prism) is split into its six faces, where each face is passed through a separate three-layer fully connected network, with a width of twice the input size (the corresponding face's size) and output size that preserves the input size. All six outputs are then combined into a single variable $\tilde{\mathscr{E}}$ such that it has the same size as $\mathscr{E}$. This prepossessing stage, which exploits the geometrical structure of the environment $\mathscr{E}$, was found to improve the overall mutual information detected by the critic. In the second stage, both $\tilde{\mathscr{E}}$ and $\mathscr{H}$ are passed through a separate three-layer connected network, with a width of twice the input size for $\tilde{\mathscr{E}}$ and fixed on 64 for the discretized $\mathscr{H}$, and fixed output size of 256 (the embedding dimension). In the third stage, the two outputs are given to the INFO-NCE estimator, which yields the lower bound on the mutual information.

The $(\mathscr{V},\mathscr{E})$ corpus was split between training and testing datasets to ensure that no over-fitting had taken place. The learning process occurred over five epochs.

\subsection{Application of RSMI-NE to the AT-TFI model}\label{sec:application_ATTFI}

The operator extraction scheme utilizes the different symmetry sectors. The $\sigma,\tau$ fields, as well as their derivatives, are odd under a global $\mathbb{Z}_2$ transformation of the spins. Thus, the $
\langle \sigma \rangle\langle \tau \rangle$ operator is odd under a $\mathbb{Z}_2\times\mathbb{Z}_2$ transformation, where each spin field can be flipped separately. Furthermore, the current operator $\langle \sigma \rangle\left<\partial\tau\right>-\langle \tau \rangle\left<\partial\sigma\right>$ is odd under both $\mathbb{Z}_2\times\mathbb{Z}_2$ and spatial inversion transformations. These facts allow us to construct the following four-stage scheme for learning the AT-TFI leading operators:
\\
1) Augment $\mathscr{V}$ such that only $\mathbb{Z}_2\times\mathbb{Z}_2$ even operators can be learned. This would first yield the operator $\langle \sigma \rangle^2-\langle \tau \rangle^2$ (scaling dimension of $\sim 1.2$) and then $\langle \sigma \rangle^2+\langle \tau \rangle^2$ (scaling dimension of $\sim 1.5$). Learn until the mutual information is exhausted.\\
2) Augment $\mathscr{V}$ such that only $\mathbb{Z}_2$ and inversion even operators can be learned. This would yield the $\langle \sigma \rangle\langle \tau \rangle$ operator (Scaling dimension of $\sim 1.2$).  Learn until the mutual information is saturated.\\
3) Finally, augment $\mathscr{V}$ such that only $\mathbb{Z}_2$ even operators can be learned. This would yield the current operator (Scaling dimension of $2.0$).\\
4) As an extra step, one can also remove any augmentation. This would yield the $\langle \sigma \rangle,\langle \tau \rangle$ operators and their derivatives.

\subsection{Application of RSMI-NE to the SD-IHG model}\label{sec:application_SDIHG}

The operator extraction scheme employed the spatial rotation and inversion symmetry sector. While the $\langle A \rangle$ and $\langle S \rangle$ operators are invariant under any spatial rotation and inversion, their derivatives, as well as the hypothesized current operators, are not. Therefore, we could augment $\mathscr{V}$ via spatial rotations and inversions such that only the $\langle A \rangle$ and $\langle S \rangle$ operators could be learned (Scaling dimensions of $\sim 1.2$ and $\sim 1.5$ respectively). Then, after the mutual information is saturated, we remove the augmentation and learn the first leading operator which is odd under the spatial symmetries.
\\
\subsection{More details on extremal configurations and projections}\label{sec:projection}

Searching for extremal configurations for the SD-IHG and AT-TFI models was performed in a gradient descent-like manner. At the start of the search, we set $\mathscr{V}$ to some initial random value within the corresponding spatial symmetry sector. Then, in each step a random d.o.f in  $\mathscr{V}$ was chosen. If flipping it decreases/increases (depending on whether we are searching for a minimal/maximal configuration) the value of the operator acting on $\mathscr{V}$ within the sector, we flip it. We repeat the latter step until an extremum is reached.

The $2D$ projections of the extremal configuration, as appear in Table \ref{tbl:SDIHG_and_ATTFI_CFT} are attained by taking a spatial slice of the extremal configuration (in the AT-TFI model, the slices are in the spatial plane). The slices are taken from the boundary of $\mathscr{V}$. In the case of the SD-IHG model, the slices were taken in the plane such that the derivative operator akin to $\partial \langle A \rangle$ was to be visually noticeable. Extremal configurations for the SD-IHG model were taken from neural operators of linear size $3$, which are qualitatively similar to the ones used for computing the scaling dimensions (with linear size $2$) but yielded less accurate results.

\begin{figure}[htp]
\centering
\adjustbox{trim={0.02\width} {.4\height} {0.45\width} {0.05\height},clip}{
\includegraphics[width=0.8\textwidth]{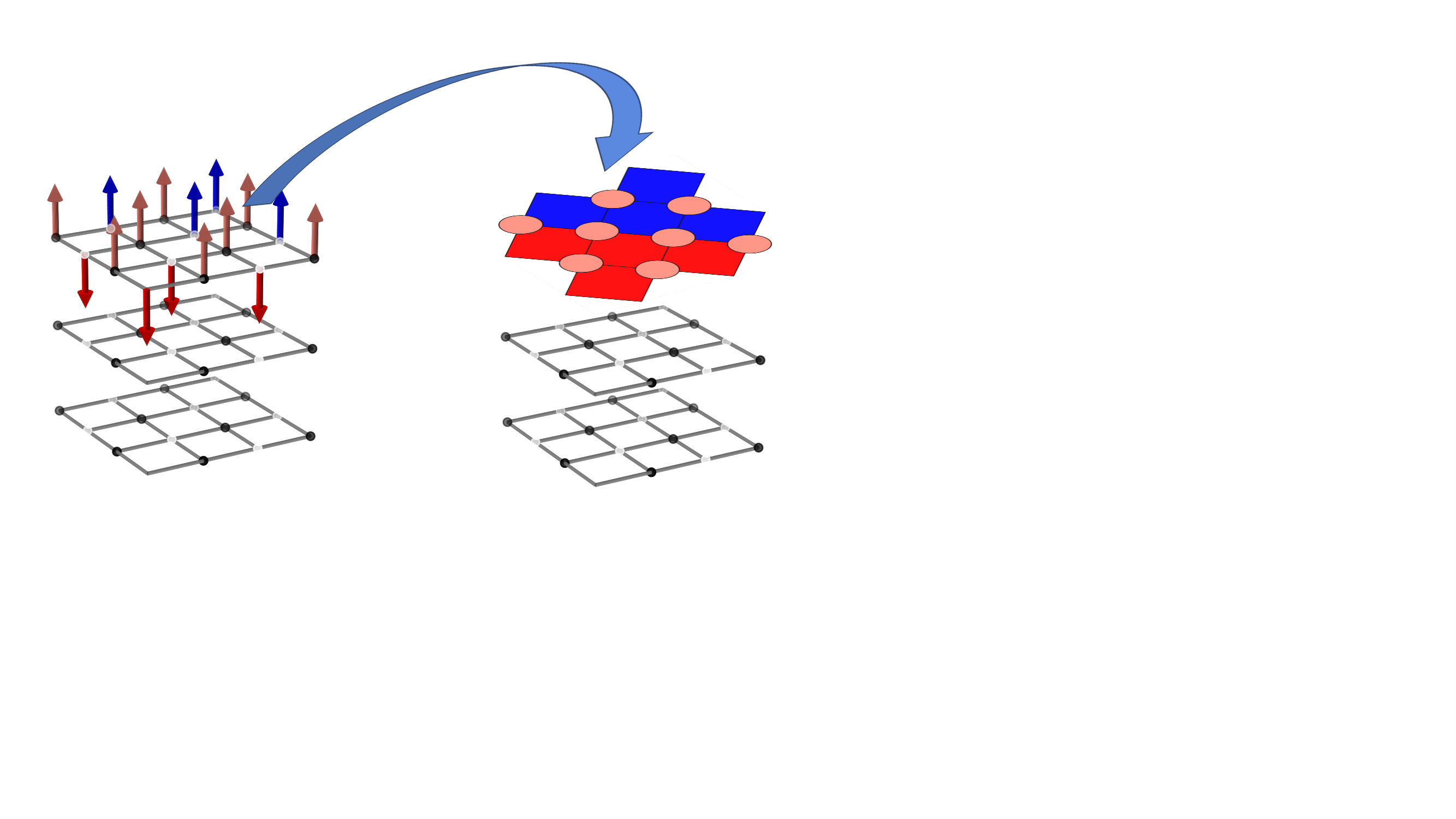}}
\caption{Visualization of an extremal projection of the current configuration for the AT-TFI model. The projection takes a spatial slice of a spins configuration; the $\tau$ sites (appear as white dots on the grid) turn into plaquettes, and the $\sigma$ sites (appear as black dots on the grid) turn into circles. The red and blue color denotes the spins' values (direction of the arrows).}
    \label{fig:ATTFI_extremal_config}
\end{figure}

The scheme for presenting the extremal projection in the case of the AT-TFI model appears in Fig. \ref{fig:ATTFI_extremal_config}.

\end{document}